%% file: ms.tex
\documentclass[12pt,preprint]{aastex}

\usepackage{natbib}
\newcommand{\kmsmpc}{{\rm km\ s\ }^{-1}{\rm\ Mpc}^{-1}}
\newcommand{\Kms}{{\rm km\ s}^{-1}}
\newcommand{\Bband}{B}
\newcommand{\Vband}{V}
\newcommand{\BminusV}{({\Bband}{\rm -}{\Vband})}

\newcommand{\Ebv}{E\BminusV}

\newcommand{\ssp}{\def\baselinestretch{1.0}\large\normalsize}
\newcommand{\gtrsi}{\mathrel{\hbox{\rlap{\hbox{\lower4pt\hbox{$\sim$}}}\hbox{$>$}}}}

\shorttitle{Cepheid Distance to NGC 1637} \shortauthors{Leonard et al.}

\begin{document}

\title{The Cepheid Distance to NGC 1637: A Direct Test of the EPM Distance to
SN 1999em}


\author{Douglas C. Leonard\altaffilmark{1},
Shashi M. Kanbur\altaffilmark{2},
Choong C. Ngeow\altaffilmark{2}, and 
Nial R. Tanvir\altaffilmark{3}
}

\altaffiltext{1}{Five College Astronomy Department, University of Massachusetts, Amherst,
 MA 01003-9305; leonard@nova.astro.umass.edu}
\altaffiltext{2}{Department of Astronomy, University of
Massachusetts, Amherst, MA 01003-9305; shashi@astro.umass.edu, ngeow@nova.astro.umass.edu}
\altaffiltext{3}{Department of Physical Science, University of Hertfordshire, College
Lane, Hatfield, AL10 9AB, UK; nrt@herts.star.ac.uk}


\begin{abstract}

Type II-plateau supernovae (SNe II-P) are the classic variety of core-collapse
events that result from isolated, massive stars with thick hydrogen envelopes
intact at the time of explosion.  Their distances are now routinely estimated
through two techniques: the expanding photosphere method (EPM), a primary
distance-determining method, and the recently developed standard-candle method
(SCM), a promising secondary technique.  Using Cycle 10 {\it Hubble Space
Telescope} ({\it HST}) observations, we identify 41 Cepheid variable stars in
NGC~1637, the host galaxy of the most thoroughly studied SN~II-P to date,
SN~1999em.  Remarkably, the Cepheid distance that we derive to NGC~1637, $D =
11.7 \pm 1.0$ Mpc, is nearly $50\%$ larger than earlier EPM distance estimates
to SN~1999em.  This is the first direct comparison between these two primary
distance determining methods for a galaxy hosting a well-observed,
spectroscopically and photometrically normal, SN~II-P.  Extensive consistency
checks show strong evidence to support the Cepheid distance scale, so we are
led to believe that either SN~1999em is in some heretofore unsuspected way an
unusual SN~II-P, or that the SN II-P distance scale must be revised.  Assuming
the latter, this one calibration yields $H_0{\rm (EPM)} = 57 \pm 15\ \kmsmpc$
and $H_0{\rm (SCM)} = 59 \pm 11\ \kmsmpc$; additional calibrating galaxies are
clearly desirable in order to test the robustness of both determinations of
$H_0$.

The {\it HST} observations of NGC~1637 also captured the fading SN~1999em two
years after explosion, providing the latest photometry ever obtained for an
SN~II-P.  The nebular-phase photometric behavior of SN~1999em closely follows
that observed for SN~1987A at similar epochs.  The $V$ and $I$ light curves are
both declining at rates significantly greater than the decay slope of
$^{56}{\rm Co}$ predicts.  This is likely due to an increasing transparency of
the envelope to gamma rays, and perhaps also to the formation of dust in the
cooling atmosphere of the supernova.  The absolute $V$-band brightness of
SN~1999em is $\sim 0.25$ mag brighter than SN~1987A at the same epochs, which
suggests that a slightly greater amount of radioactive $^{56}{\rm Ni}$, $\sim
0.09\ M_\odot$, was ejected by SN~1999em than was derived for SN~1987A ($0.075\
M_\odot$).

\end{abstract}


\section{INTRODUCTION}
\label{sec:introduction}
Direct comparisons between distances estimated by two {\it primary}
extragalactic distance indicators are rare. The empirical period-luminosity
(PL) relation of Cepheid variable stars remains the most widely used and well
studied primary extragalactic distance-determining method.  It currently
provides our best distance estimates to over thirty nearby ($D \lesssim 23$
Mpc) galaxies, which enable the calibration of a host of secondary distance
indicators that then establish galaxy distances well into the Hubble flow.
Using such measurements, three groups have independently estimated Hubble's
constant with total (statistical + systematic) reported
uncertainties\footnote{In this paper we abide by the following error bar
conventions: a single, unqualified error bar (i.e., not specifically labeled as
``statistical'' or ``systematic'') represents the total (statistical and
systematic, added in quadrature) uncertainty of the measurement; if two
uncertainties are given, the first represents the statistical, and the second
the systematic, error of the measurement.}  of only $\sim 10\%$: $H_0 = 58 \pm
6\ \kmsmpc$ (\citealt{Tammann02a}, hereafter T02), $H_0 = 67 \pm 7\ \kmsmpc$
\citep*{Tanvir99c}, and $H_0 = 72 \pm 8\ \kmsmpc$ \citep[][hereafter
F01]{Freedman01}.  The differences among the Hubble constants derived by the
groups are a cause of concern, and may belie unaccounted systematic
uncertainties arising from the calibration of the secondary distance indicators
(see F01, T02, \citealt{Tammann02b}, and \citealt{Gibson00} for further details
about the controversy).

The Cepheid-based distance scale has recently passed a stringent external test
of its accuracy: for NGC~4258, the $H_0$ Key Project (hereafter KP) distance of
$7.8 \pm 0.3 \pm 0.5$ Mpc \citep{Newman01} agrees well with the geometric
distance determined from observations of line-emitting water masers orbiting a
supermassive black hole at the galaxy's center, of $7.2 \pm 0.5$ Mpc
\citep{Herrnstein99}.  Since the maser distance to NGC~4258 is believed to have
few potential sources of unaccounted systematic error, it provides a robust
external check on the accuracy of the Cepheid distance scale.  The agreement
between the Cepheid and maser distances to NGC~4258 builds confidence in the
Cepheid PL relation as an accurate primary extragalactic distance indicator.

The extraordinary brightness of supernovae (SNe) has long made them attractive
targets for extragalactic distance measurements as well.  SNe are grouped into
two main categories, based on whether the supernova (SN) results from the
thermonuclear explosion of a white dwarf (Type~Ia) or the core collapse and
subsequent envelope ejection of a massive (initial mass $\gtrsi 8-10\
M_{\odot}$; see \citealt{Woosley86}, and references therein) star (Types Ib,
Ic, and II; see \citealt{Filippenko97} for a general review of SN types).
Recently, the standard-candle assumption for SNe Ia, coupled with empirical
relations to correct ``non-standard'' events \citep*{Riess96,Phillips93}, has
emerged as the leading secondary distance-determination method.  SNe~Ia provide
precise relative distances for dozens of SNe Ia at $z \gtrsim 0.3$, which have
revealed evidence for a nonzero cosmological constant
\citep{Riess98,Perlmutter99}.

Unlike SNe Ia, SNe II have not traditionally served as secondary distance
indicators through a standard-candle assumption.  Rather, through calibration
with the expanding photosphere method (EPM; \citealt{Kirshner74}), a variant of
the \citet{Baade26} method used to determine the distance to variable stars,
they have served as {\it primary} distance indicators, providing direct
distance measurements completely independent of the Cepheid scale.  ``Normal''
core-collapse SNe are thought to result from isolated, massive stars with thick
hydrogen envelopes intact at the time of the explosion.  Their light curves
show a distinct $V$-band plateau (hence the subclassification, ``SN II-P''),
resulting from an enduring period ($\sim 100$ days) of nearly constant
luminosity as the hydrogen recombination wave recedes through the envelope and
slowly releases the energy deposited by the shock and by radioactive decay.
EPM in its most general form (see \S~\ref{sec:tepmatdtsn}) is {\it only}
securely applied to SNe~II-P, and not to ``peculiar'' core-collapse events such
as those of Type II-L (``L'' for their linearly declining optical light curves,
lacking a $V$-band plateau; these events are believed to result from
progenitors that have lost a substantial fraction of their hydrogen envelope
prior to exploding) or those resembling SN~1987A, which showed spectral and
photometric peculiarities thought to be due to the compact nature of its
progenitor star \citep{Woosley87}.  From a consistently applied EPM analysis of
nine normal SNe II-P spanning distances out to $\sim 200$ Mpc, a value of $H_0
{\rm (EPM)}\ = 71 \pm 9\ {\rm(statistical)}\ \kmsmpc$ has recently been derived
(\citealt{Hamuy01a}; Hubble constant obtained by averaging the six values for
$H_0$ derived by using the $BV$, $BVI$, and $VI$ passbands to estimate
photospheric color-temperature and two different techniques to estimate
photospheric velocity for each filter combination; see
\S~\ref{sec:tepmatdtsn}).

While the statistical uncertainty in Hubble's constant derived from EPM
distances is straightforward to derive from the scatter of the EPM distances in
the Hubble diagram, the external accuracy of the technique is rather hotly
debated.  Fueling the debate is the fact that there have never been any
measurements of Cepheid variables in a galaxy that has hosted a well-observed,
spectroscopically and photometrically normal, SN II-P.  To be sure, Cepheid
distances do exist to four galaxies that have hosted non-plateau Type II
events, as well as to one poorly observed SN~II-P.\footnote{The four
non-SNe~II-P are SN~1970G in NGC~5457 and SN~1979C in NGC~4321 \citep*[both
Type~II-L;][]{Schmidt92}, SN~1987A in the Large Magellanic Cloud (LMC), and
SN~1989L (classified as a Type~IIn by \citealt{Schlegel90}, a Type~II-L by
\citealt{Patat93,Patat94}, and a Type~II-P with spectroscopic peculiarities by
\citealt{Schmidt94}).  A Cepheid distance exists to NGC~3627, the host of the
Type II-P SN~1973R.  Unfortunately, SN~1973R was not well observed, and the EPM
distance has a statistical uncertainty of nearly $50\%$ \citep*[][hereafter
E96]{Eastman96}.}  However, the non-standard nature of these SNe, or their
poorly estimated distances, prevent strong conclusions from being drawn
\citep{Leonard5}.

SN 1999em, a bright, extremely well-observed SN II-P in the nearby SBc galaxy
NGC 1637, presents an outstanding opportunity to finally directly test the
consistency between these two primary distance techniques.  It is by far the
best studied normal SN~II-P to date, with three independent and mutually
consistent EPM distances now reported: $7.50 \pm 0.50 {\rm\ (statistical)\ }$
Mpc \citep{Hamuy01b}, $8.24 \pm 0.56$ (statistical) Mpc \citep{Leonard5}, and
$7.83 \pm 0.30$ (statistical) Mpc \citep{Elmhamdi03}.  (A subsequent reanalysis
of the \citealt{Hamuy01b} data is given by \citealt{Hamuy01a}, in which 32
different EPM distances to SN~1999em are compared using slightly different
implementations of the technique; a single ``best'' result, however, is not
given.)  In this paper, we present the analysis of multi-epoch Wide Field
Planetary Camera 2 (WFPC2) Cycle 10 {\it Hubble Space Telescope} ({\it HST})
observations of NGC~1637, which were taken in order to derive its distance
through the analysis of its Cepheid variable stars.  As a bonus, photometry of
SN~1999em itself is obtained from these images, permitting a study of the
photometric behavior of an SN~II-P two years after explosion.  This is the
latest photometry ever obtained for this important class of core-collapse SN.

The paper is organized as follows.  We review the Cepheid PL relation and the
EPM technique in \S~\ref{sec:preliminaries}, and present the {\it HST}
observations of NGC~1637 along with a description of the data reduction and
photometric analysis in \S~\ref{sec:observationsandphotometry}.  We identify
the Cepheid variables, derive the Cepheid distance to NGC 1637, and provide a
brief analysis of the late-time photometric behavior of SN~1999em in
\S~\ref{sec:results}.  We compare the EPM and Cepheid distances to NGC~1637,
and use the Cepheid calibration to derive Hubble's constant from both the EPM
and the ``standard-candle method'' (SCM) recently developed by \citet{Hamuy02}
for SNe~II-P, in \S~\ref{sec:discussion}. We summarize our conclusions in
\S~\ref{sec:conclusions}.  Throughout the paper, frequent reference is made to
the work of the two main groups that have previously studied the Cepheid
distance scale using {\it HST}: The KP, led by W. Freedman, and the Type Ia SNe
{\it HST} Calibration Program, led by A. Saha, G. Tammann, and A. Sandage
(hereafter STS). The most recent summary papers from each group, F01 and T02
(for KP and STS, respectively), are cited often.

\section{Deriving Distances to Cepheids and Type II-Plateau Supernovae}
\label{sec:preliminaries}
Before presenting the Cepheid study of NGC~1637, we first briefly review the
techniques used to measure distances to Cepheid variables through the PL
relation and SNe II-P through the EPM, focusing particular attention on
potential sources of {\it systematic} uncertainty in both techniques.

\subsection{Cepheid Variables and the Period-Luminosity Relation}
\label{sec:cvatplr}

In principle, once a Cepheid variable star in a target galaxy has been
identified and had its period determined and mean magnitudes measured in at
least two photometric bands, estimating the distance to its host galaxy is
straightforward.  The fundamental equation underlying the technique is the
empirical PL relation
\begin{equation}
M_C = a_C \log(P) + b_C,
\label{eqn:one}
\end{equation}
\noindent where $M_C$ is the intensity mean absolute magnitude of a Cepheid in
photometric band $C$, $P$ its period (usually measured in days), and $a_C$ and
$b_C$ the slope and zero-point, respectively, of the relation.  The physics
behind the correlation is well understood as a consequence of the period--mean
density relation obeyed by pulsating stars in the region of the
Hertzsprung-Russell diagram, the instability strip, where variable stars are
observed \citep{Sandage58,Madore91}.  Since the LMC is nearby and thought to
have little depth along the line of sight and comparatively low extinction, PL
relations have traditionally been established through the examination of its
Cepheids, although attempts at a Galactic PL relation have recently been made
(see discussion below and in \S~\ref{sec:discussion1}).  Since our {\it HST}
study relies exclusively on $V$ and $I$ photometry, we consider only these
bandpasses in what follows.

Upon adopting a set of PL relations and computing $M_V$ and $M_I$ through
equation~(\ref{eqn:one}), a Cepheid's {\it apparent} distance moduli are
derived from its observed mean magnitudes, $m_V$ and $m_I$, as
\begin{equation}
 \begin{array}{lll}
  \mu_V  & \equiv & m_V - M_V,\\
  \mu_I  & \equiv & m_I - M_I.
 \end{array}
\label{eqn:two}
\end{equation}
\noindent These apparent distance moduli are then duly corrected for
extinction,
\begin{equation}
 \mu_W \equiv \mu_V - A_V = \mu_I - A_I,\\
\label{eqn:four}
\end{equation}

\noindent where $A_V$ and $A_I$ represent the extinction in the $V$ and $I$
bands, respectively, and the subscript $W$ refers to the ``Wesenheit
reddening-free index'' \citep{Madore82,Tanvir97}, which for $VI$ photometry is
defined as $m_W \equiv m_V - R(m_V - m_I)$; this index is discussed in more
detail later in this section.  From equation~(\ref{eqn:four}) we see that
\begin{equation}
\mu_V - \mu_I = A_V - A_I,
\label{eqn:five}
\end{equation}
\noindent which simply expresses the fact that, absent measurement or other
systematic errors (see discussion by \citealt*{Saha00}), $\mu_V$ will exceed
$\mu_I$ by the difference between the extinction in the two passbands.
Equations~(\ref{eqn:two}), (\ref{eqn:four}), and (\ref{eqn:five}) are then
combined with the relation
\begin{equation}
R \equiv \frac{A_V}{A_V - A_I},
\end{equation}
\noindent where $R$ is typically taken to be $2.45$ (derived from the
extinction curve of \citealt*{Cardelli89}), to produce the unreddened distance
modulus of the Cepheid:
\begin{equation}
\mu_W = \mu_V - R(\mu_V - \mu_I).
\label{eqn:eight}
\end{equation}
The unweighted mean, $\overline{\mu_W}$, of all of the individually determined
Cepheid distance moduli (typically a few dozen for an {\it HST} target galaxy)
then provides the distance modulus of the galaxy (uncorrected for possible
metallicity effects, see discussion below).  The use of a weighting scheme to
combine the individual Cepheid distance moduli is avoided since quantifiable
photometric errors are believed to be less significant than the effects that
the width of the intrinsic PL relation and other systematics have on the
individual distances.

For ease of analysis, equation~(\ref{eqn:one}) is often written explicitly in
terms of the equivalent reddening-free index, $W$, such that
\begin{equation}
M_W \equiv a_W \log(P) + b_W,
\label{eqn:nine}
\end{equation}
\noindent and 
\begin{equation}
 \begin{array}{lll}
   \mu_W & \equiv & m_W - M_W\\
         & =      & m_W - (a_W \log[P] + b_W),\\
 \end{array}
\label{eqn:ten}
\end{equation}
\noindent where $M_W = M_V - R(M_V - M_I), m_W = m_V - R(m_V - m_I),
a_W = a_V - R(a_V - a_I)$, and $b_W = b_V - R(b_V -
b_I)$. Equation~(\ref{eqn:ten}) is formally identical to
equation~(\ref{eqn:eight}).  The advantage of the Wesenheit approach
is that the ``reddening corrected'' PL relation (i.e., $\log (P)$
vs. $m_W$) for a sample of Cepheids is intrinsically tighter than the
PL relations in either band individually, and, at least in principle,
permits analysis without the complicating effects of extinction. 

Using {\it HST}, the preceding technique (or one closely similar to it) has
been applied to Cepheid variables (generally with $10 < P < 60$ days) in over
30 galaxies, providing distances to each galaxy with typical {\it statistical}
uncertainties (e.g., $\sigma[m_W] / \sqrt{n-1}$, where $n$ is the number of
Cepheids identified in the galaxy) of $\lesssim 5\%$.  Despite intense effort
during the past decade, however, significant {\it systematic} uncertainty still
exists, and is important to quantify; detailed discussions are provided by
\citet{Tammann02b}, T02 and F01.  For our Cepheid distance determination to
NGC~1637, we identify the following five sources of systematic uncertainty:

1. {\it Zero point of the PL relation.}  The LMC distance modulus currently
sets the zero-point of the PL relations, and therefore has a direct effect on
inferred Cepheid distances to other galaxies.  By considering dozens of
independent distance estimates to the LMC, F01 adopt a value of $\mu_{\rm LMC}
= 18.50 \pm 0.1$ mag ($50.1 \pm 2.3$ kpc).  Using a similar but somewhat more
selective approach, T02 derive $\mu_{\rm LMC} = 18.56 \pm 0.02$ mag.  We
conservatively adopt 0.1 mag for the zero-point uncertainty, and select
$\mu_{\rm LMC} = 18.50$ mag as the distance modulus of the LMC.

2. {\it Metallicity.}  Empirical and theoretical evidence suggests that the
   Cepheid PL relation has a small dependence on metallicity
   (\citealt{Kennicutt98,Sasselov97,Kochanek97}; \citealt*{Bono00}).  Such a dependence
   necessitates the use of a ``metallicity correction term'', $\delta \mu_Z$,
   which is dependent on the metallicity difference between the target galaxy
   and the galaxy used to derive the PL relations (e.g., the LMC), to correct
   the unreddened distance modulus, $\mu_W$ (eq.~[\ref{eqn:ten}]).  This is
   accomplished by deriving the ``true'' distance modulus, $\mu_0$, as
\begin{equation}
\mu_0 = \mu_W + \delta\mu_Z. \\
\label{eqn:twelve}
\end{equation}
\noindent The value of the correction term is still uncertain.  Ultimately,
F01 adopt\footnote{Note that the metallicity correction equation given by F01,
$\delta\mu_Z = (-0.2 \pm 0.2)([{\rm O/H}]_{\rm galaxy} - [{\rm O/H}]_{\rm
LMC})$, evidently contains a sign error, since the metallicity corrections
derived by F01 for individual galaxies are consistent with
equation~(\ref{eqn:thirteen}) of the present paper.}:
\begin{equation}
\delta\mu_Z = (0.2 \pm 0.2)([{\rm O/H}]_{\rm galaxy} - [{\rm    O/H}]_{\rm
LMC}),\\
\label{eqn:thirteen}
\end{equation}
\noindent where $[{\rm O/H}] \equiv 12 + \log{\rm O/H}$.  The uncertainty
contributed to the final distance is therefore a direct function of the
metallicity difference between the target galaxy and the LMC. The metallicity
of a galaxy, $[{\rm O/H}]_{\rm galaxy}$, is typically determined from
\ion{H}{2} region studies.  For the LMC, $[{\rm O/H}]_{\rm LMC} = 8.50$.  For
NGC 1637, the mean metallicity of the 15 \ion{H}{2} regions studied by
\citet{Vanzee98} yields $[{\rm O/H}]_{\rm NGC1637} = 9.08$.  (The
\citeauthor{Vanzee98} study is also consistent with the recent detailed study
of a single \ion{H}{2} region in NGC~1637 by \citealt*{Castellanos02}.)  The
higher metallicity of NGC~1637 compared with the LMC implies a correction to
its unreddened Cepheid distance of $\delta\mu_Z = 0.12 \pm 0.12$ mag, which we
adopt.

3. {\it Slope of the PL relations.}  The values of $a_V$ and $a_I$ in
   equation~(\ref{eqn:one}) are the subjects of ongoing refinement and
   controversy, and represent an area of disagreement between the methodologies
   of the KP and STS groups.  Table~\ref{tab:tab1} lists four determinations of
   their values, and the zero points, $b_V$ and $b_I$, that result from
   assuming $\mu_{\rm LMC} = 18.50$ mag for the relations derived from LMC
   Cepheids (note that in the latest STS paper by \citealt{Thim03}, an LMC
   distance modulus of 18.50 mag is adopted).  The Cepheid samples used to
   derive the four PL relations are as follows.

\noindent{\it Set A:} Photometry of $\sim 650$ LMC Cepheids with periods
ranging from 2.5 to 31 days that were observed as part of the Optical
Gravitational Lensing Experiment (OGLE; \citealt{Udalski99}).  The PL relations
are those of F01.

\noindent{\it Set B:} A sample of 32 Cepheids in the LMC with $1.6 < P < 63$
days, presented by \citet{Madore91}.  

\noindent{\it Set C:} The OGLE LMC Cepheid sample from Set {\it A} (above), but
restricted to the 44 Cepheids with $P > 10$ days.  The PL relations are those
of \citet{Thim03}.

\noindent{\it Set D:} 53 Galactic Cepheids drawn from the samples of
\citet{Feast99} and \citet*{Gieren98}.  The PL relations are those of
\citet{Thim03}.

In the final KP paper, F01 adopt the PL relations defined by Set {\it A}, and
apply metallicity corrections as given by equations~(\ref{eqn:twelve}) and
(\ref{eqn:thirteen}).  In the most recent paper from the STS group,
\citet{Thim03} derive the distance to NGC~5236 by taking the {\it average} of
the distances found by applying the relations defined by Sets {\it B, C,} and
{\it D} individually, with no metallicity correction.  Thorough discussions of
the relative merits of different PL relations are given by \citet{Tanvir99a}
and \citet{Kanbur03}.  For our present study, we shall follow the KP approach,
and use the PL relations defined by Set {\it A} to determine the distance to
NGC~1637.  However, to help gauge the influence that the adopted PL relation
makes on our derived distance, we shall also determine the distance to NGC~1637
using the STS methodology for comparison (see \S~\ref{sec:discussion1}).

Recent assessments of the systematic uncertainty contributed by the uncertainty
in the slope of the PL relation are given by \citet{Tammann02b}, who find that
the slope uncertainty may contribute as much as a $6\%$ ($0.13$ mag) systematic
uncertainty to the Cepheid distance scale, and \citet{Sebo02}, who posit an
$\sim0.06$ mag maximum contribution.  Since much of the slope differences may
in fact be caused by metallicity differences \citep{Tammann02c}, for which a
separate contribution has already been included, we adopt the smaller, $0.06$
mag, estimate for the uncertainty; we shall show in \S~\ref{sec:discussion1}
that, at least for NGC~1637, this appears to be an appropriate uncertainty
estimate.

4. {\it Aperture corrections.}  All photometric measurements in our study were
   made using the HSTphot stellar photometry package \citep{Dolphin00a}, and
   are detailed in \S~\ref{sec:thwpp}.  Since the NGC~1637 field lacks bright,
   isolated stars from which to produce empirical aperture corrections, we
   adopt the default aperture corrections for the F555W and F814W filters, for
   which the HSTphot manual\footnote{HSTphot user manual available at
   \url{http://www.noao.edu/staff/dolphin/hstphot/}.} assigns an uncertainty of
   $0.02$ mag.

5. {\it WFPC2 zero point.}  We adopt the estimate by \citet{Parodi00} of $0.04$
   mag for the uncertainty in the WFPC2 zero point.

   The sources of systematic uncertainty are summarized in
   Table~\ref{tab:tab2}.  Adding them together in quadrature yields a total
   systematic uncertainty for the Cepheid technique of $0.17$ mag, or somewhat
   less than $10\%$, for our Cepheid distance estimate to NGC~1637.  

\subsection{The Expanding Photosphere Method}
\label{sec:tepmatdtsn}

\subsubsection{The EPM Technique}
\label{sec:tet}

Since a thorough review of EPM is given by \citet{Leonard5}, we provide only a
brief summary of the method here.  In essence, EPM is a geometrical technique:
the linear radius of the expanding SN photosphere, $R$, is compared with its
angular size, $\theta$, to derive the distance to the SN, $D$.  Naturally,
since all extragalactic SNe are unresolved during the ``photospheric'' phase
(the plateau in an SN~II-P), $R$ and $\theta$ must be derived rather than
measured directly.  Briefly, the radial velocity of the expanding photosphere,
$v$, is found from the Doppler shifting of the spectral lines
\citet{Kirshner74}, so that $R = v(t - t_0)$, where ($t - t_0$) is the time
since explosion and the SN is assumed to be in free expansion.  The
photosphere's theoretical angular size, $\theta$, is calculated by comparing
the observed flux with that predicted from theoretical models (e.g., a
``dilute'' blackbody; \citealt{Wagoner82}) for a spherical SN photosphere as
\begin{equation}
\theta = \sqrt{\frac{f_\nu 10^{0.4A_\nu}}{\zeta_\nu^2 (T_c) \pi B_\nu (T_c)}} ,
\label{eqn:fourteen}
\end{equation}
\noindent where $B$ is the Planck function at color temperature $T_c$, $f$ is
the flux density received at Earth, $A$ is the extinction, and $\zeta(T_c)$ is
the theoretically derived ``dilution factor.''  Because accurate
spectrophotometry is generally not available, equation~(\ref{eqn:fourteen}) is
typically recast in terms of broadband photometry, with $T_c$ and $\zeta$
derived for some subset of $BVIJHK$.  Since the spectral and photometric data
are usually not coincident with each other, an interpolation of either the
photometry to the epochs of the spectral observations (the preferred technique
when many spectra with good temporal sampling exist) or the photospheric
velocity measurements derived from the spectra to the times of the photometry
(used when few spectra are available, which is more typically the case) must be
made.

With $R$ and $\theta$ known, $D$ can be found since $\theta = R/D$ in the
small-angle approximation.  The greater the number of
observational epochs (a minimum of 2 is required), the more precise $D$
becomes; in essence, each observational epoch of the SN provides an independent
estimate of its distance.

\subsubsection{Sources of Statistical Uncertainty}
\label{sec:ssu}

By analyzing the scatter in the EPM Hubble diagram, \citet{Hamuy01a} finds that
the internal consistency among EPM distances indicates an average statistical
uncertainty of $\sim 20\%$ for an EPM distance measurement, a value that
implicitly includes all of the contributions from {\it statistical} errors in
photometry, photospheric velocity measurements, extinction, dilution factors,
and peculiar motions of the host galaxies.  This value also necessarily
includes any effects resulting from departures from sphericity of SN~II-P
photospheres which would lead to a directionally dependent luminosity and,
hence, derived EPM distance.  This leads \citet{Hamuy01a} to conclude that such
departures must on average be small, since the other sources, which are more
easily quantified, can easily account for a $20\%$ statistical uncertainty
without needing to include a contribution from asphericity.  This contention is
consistent with spectropolarimetric studies of core-collapse SNe, which also
find weak evidence for asphericity during the plateau phase for SNe~II-P
(polarimetry directly probes asphericity since in general the greater the
polarization, the greater the implied asphericity; see, e.g.,
\citealt*{Leonard1}; \citealt{Leonard4}).  In fact, the most thorough
spectropolarimetric investigation of the sphericity assumption for an SN~II-P
is for SN~1999em itself, for which \citet{Leonard3} report a very low intrinsic
polarization which, for most viewing orientations and plausible geometries,
suggests a substantially spherical electron-scattering atmosphere
\citep[e.g.,][]{Hoflich91}.  Ongoing spectropolarimetric studies of SNe~II-P
continue to support the contention that asphericity is not a major concern for
the EPM technique.

A typical statistical uncertainty in an EPM distance measurement is therefore
$\sim 20\%$.  For an extremely well observed SN~II-P like SN~1999em,
quantifiable sources of statistical uncertainty may be even smaller, as
suggested by the $\sim 10\%$ uncertainties assigned by \citet{Hamuy01b},
\citet{Leonard5}, and \citet{Elmhamdi03} for their EPM distance estimates of
this event.

\subsubsection{Sources of Systematic Uncertainty}
\label{sec:tdf}

While the internal precision of the EPM is fairly straightforward to constrain,
the external accuracy of the technique is not.  Unlike the case for Cepheids
(\S~\ref{sec:introduction}), no rigorous external check of the accuracy of the
EPM technique for an SN~II-P has been carried out prior to the present study.
It is generally agreed that the largest potential source of systematic error in
applying the EPM to SNe~II-P is the theoretically derived dilution factor,
$\zeta(T_c)$, discussed in the previous section.  Accurate knowledge of
$\zeta(T_c)$ is of paramount importance since it directly impacts derived
distances, in some cases by over a factor of two; in fact, it is often referred
to as the ``distance correction factor,'' since it ``corrects'' EPM-derived
distances such that
\begin{equation}
D_{\rm actual} = \zeta D_{\rm measured},
\label{eqn:fifteen}
\end{equation}
\noindent where $D_{\rm measured}$ is the distance derived without its
inclusion in equation~(\ref{eqn:fourteen}).  In principle, the dilution factor
could depend on many things, including the chemical composition (i.e., the
metallicity) and density structure of the progenitor star, and the expansion
rate and luminosity of the SN explosion.  However, studies of theoretical
models of realistic SN atmospheres with a wide range of properties have
demonstrated that $\zeta$ is in fact a nearly one-dimensional function of color
temperature, $T_c$ (E96; \citealt{Montes95}).  In principle, this simple
dependence on an observable quantity, $T_c$, allows average dilution factors to
be derived from theoretical models that may then be broadly applied to {\it
all} normal SNe~II-P, and obviates the need to custom-craft individual models
for particular events.  Currently, the only published values of $\zeta(T_c)$
for SNe II-P are those of E96, and so they have been used exclusively in EPM
studies to date.  It is important to point out again that the ``generic''
dilution factors of E96 are only applicable to SNe~II-P, and not to
``peculiar'' core-collapse events, such as those in which the progenitor star
is thought to have experienced substantial mass loss prior to exploding (e.g.,
SNe~II-L, IIb, IIn, Ib, Ic), or SN~1987A-like events.  While ambitious attempts
have been made to apply the general dilution factors to such events
\citep[e.g.,][]{Schmidt92,Schmidt93,Hamuy01a}, E96 stress that such peculiar
objects require individual, detailed modeling (see, e.g., \citealt{Eastman89}
for detailed modeling of SN~1987A) to derive the appropriate dilution factors.

The flash point of controversy in the EPM is the fact that all modern EPM
applications to normal SNe~II-P have relied on dilution factors produced by
only one modeling group, those published by E96 (the E96 dilution factors are
based on the radiative transfer code EDDINGTON; \citealt{Eastman93}).  (The
dilution factors of E96 have recently been slightly modified by
\citealt{Hamuy01b} to include the effects of terrestrial atmospheric absorption
and to incorporate improvements in the broadband filter functions.)  The values
produced by this group, however, have been criticized by other modelers (e.g.,
\citealt{Baron95}; \citealt{Schmutz90}).  Quantifying the degree of systematic
uncertainty in the dilution factors (the E96 models show an internal
statistical scatter of only $\sim 10\%$) is difficult, however, since no other
independent modeling group has thus far published average dilution factors
directly applicable to SNe II-P.

Some insight into the situation may be gained by considering the two {\it
unusual} core-collapse events (i.e., not normal SNe~II-P) for which independent
dilution factors from another group have been derived: SN~1987A in the LMC
\citep{Mitchell02} and SN~1993J, a ``Type~IIb'' event \citep{Filippenko88} in
M81 (\citealt{Baron95}); see \citet*{Baron94} for a description of the
radiative transfer code, PHOENIX, used in both of these calculations.  Since
\citet{Leonard6} provide a thorough review, we just quote the result: the
dilution factors produced by the PHOENIX calculations are $\sim 50\% - 60\%$
greater than those of E96.  Of course, since E96 stress that their dilution
factors are only appropriate for SNe II-P, and not to peculiar variants of the
SN~II subclass, it is not clear how meaningful the comparison is.

Taken without consideration of the peculiar nature of the two SNe~II for which
they were derived, though, equation~(\ref{eqn:fifteen}) implies that distances
obtained using the PHOENIX-derived dilution factors would be $\sim 50\% - 60\%$
greater than those that rely on the values given by E96.  Indeed, application
of the {\it generic} EPM prescription, using the E96 dilution factors, do yield
distance estimates to these two events that are significantly shorter than
those determined by other methods.  For SN~1993J, \citet{Schmidt93} derive
$D_{\rm EPM} = 2.6 \pm 0.4$ Mpc, which is about $40\%$ shorter than the Cepheid
value for M81, $3.63 \pm 0.13$ Mpc (F01).  For SN~1987A, \citet{Hamuy01a}
derives distances ranging from 28 to 39 kpc depending on the filter set used to
measure $T_c$.  These LMC distances are significantly shorter than the accepted
distance of $\sim 50$ kpc (\S~\ref{sec:cvatplr}).  While these comparisons are
suggestive, an independently derived set of dilution factors specifically
constructed for normal SNe~II-P is clearly needed, so that a direct comparison
with the results of E96 can be made.  We shall return to this contentious
issue, and discuss current work that seeks to rectify the situation, in
\S~\ref{sec:discussion3}.

A second potential source of systematic error arises from differences of
opinion regarding the best way to estimate photospheric velocities from the
spectra and to then interpolate them to derive values that are simultaneous
with the photometry (from which the color temperature and, hence, $\zeta$ are
derived).  Sparing the details (see \citealt{Hamuy01a}, \citealt{Hamuy01b}, and
\citealt{Leonard5} for extensive discussions), the bottom line is that the
usage of different techniques and interpolation schemes can lead to significant
distance discrepancies for an individual SN, in some cases of over $30\%$
\citep{Hamuy01a}. It is not at all clear at the present time which of the
various techniques is the most appropriate one to apply, and, for now, it seems
that different investigators will continue to use somewhat different methods.
Indeed, \citet{Hamuy01b}, \citet{Leonard5}, and \citet{Elmhamdi03} each use a
different technique to estimate photospheric velocity in their EPM analyses of
SN~1999em.  \citet{Hamuy01a} introduces yet another variation in the
photospheric velocity determination, this time involving the interpolation
scheme.  Clearly, thorough theoretical and empirical investigations of the
various techniques used to derive and interpolate photospheric velocities are
needed to help clarify this potentially large source of systematic error.

Between the dilution factor and photospheric velocity, then, we must allow that
the EPM distance scale may contain significant sources of systematic
uncertainty, perhaps even as large as $\sim 60\%$.  In light of the potentially
large systematic uncertainty in the EPM technique compared with Cepheids ($\sim
10\%$), it is clear that our Cepheid/EPM distance comparison for NGC~1637 will
serve primarily as an external check on the EPM-based distance scale, and not
the Cepheid scale.

\subsubsection{The EPM Distance to SN 1999em}
\label{sec:tepmdtsnn}

Discovered within six days of the explosion and achieving a peak optical
magnitude of $m_V \approx 13.8$ mag \citep{Leonard5}, SN~1999em has taken its
place as the most thoroughly studied SN~II-P to date, and has been followed
extensively at X-ray, UV, optical, IR, and radio wavelengths (see, e.g.,
\citealt{Pooley02,Leonard5,Hamuy01b,Baron00,Elmhamdi03}; \citealt*{Berger02}).
Using independent data sets, three groups, all using the dilution factors of
E96 (as modified slightly by \citealt{Hamuy01b}), have derived mutually
consistent EPM distances to SN~1999em (\S~\ref{sec:introduction}).  Taking the
simple mean of the three values yields
\begin{eqnarray*}
  \mu_0 & = & 29.48 \pm 0.14  {\rm\ (statistical)\ mag,\ or}\\
   D_{\rm EPM} & = & 7.86 \pm 0.50\ {\rm (statistical)\ Mpc}, 
\end{eqnarray*}
\noindent
where the estimated statistical uncertainty approximates the statistical
uncertainties reported for each of the three measurements, and also encompasses
the internal $1\ \sigma$ spread among the three individual measurements
themselves.  We adopt this as the ``best'' EPM distance estimate to SN~1999em.
Although the estimated statistical uncertainty in the EPM distance is rather
small, the ongoing controversies surrounding the theoretically derived dilution
factors and the technique to estimate and interpolate photospheric velocity
(\S~\ref{sec:tdf}) remain elusive gremlins in the EPM machinery, and caution us
that the total uncertainty on the EPM distance may be very large.

\section{OBSERVATIONS AND PHOTOMETRY}
\label{sec:observationsandphotometry}

\subsection{{\it HST} Observations of NGC~1637}
\label{sec:oonaip}

We obtained repeated images of a field containing nearly the entire visible
region of NGC~1637 using the WFPC2 \citep{Holtzman95} on the {\it HST} over a
60 day period from 2001 September 02 to October 31 ({\it HST} observing program
GO-9155).  A composite mosaic image including all four chips of the instrument
is shown in Figure~\ref{fig:1}.  We took observations at 12 discrete epochs in
the F555W passband and at six epochs in the F814W passband during the 60 day
window.  To facilitate removal of cosmic rays, each epoch in each filter
consisted of two successive 1100 s exposures taken during one orbit.  A journal
of observations is given in Table~\ref{tab:tab3}.

The spacing of the observations was selected using a power-law time series in order
to minimize period aliasing and maximize uniformity of phase coverage for the
expected range of Cepheid periods, $10-60$ days \citep[e.g.,][]{Freedman94}.
All observations were taken at the same telescope pointing and roll angle,
which greatly facilitated the subsequent photometry and Cepheid identification.
The frames were preprocessed through the standard Space Telescope Science
Institute pipeline using the latest calibrations as of 2002 July 19.

\subsection{The HSTphot WFPC2 Photometry Package}
\label{sec:thwpp}

\subsubsection{Image Processing}
\label{sec:ip}

The images were further processed using the suite of programs designed
specifically for the reduction of WFPC2 data that are available as part of the
HSTphot \citep{Dolphin00a} software package (version 1.1; our implementation
includes all updates through 2002 November 6).  (Note that we adopt the
convention of italicizing the names of specific tasks contained within the
HSTphot package in what follows; this convention also applies to the {\it
hstphot} task that is contained within the HSTphot package.) We first masked
bad pixels identified by the data quality images with the task {\it mask}, then
removed cosmic rays with {\it crmask}, and finally combined each epoch's
cosmic-ray-split frames with {\it coadd}.  Next, we ran {\it getsky} on each
combined frame, which creates a ``sky'' frame in which the original counts in
every pixel are replaced by a robust mean of the values contained in a
bordered square centered on the pixel.  Using these ``sky'' frames, the {\it
hotpixels} task was then used to identify and mask hot pixels and to produce
the final frames that were sent to the {\it hstphot} task for photometric
analysis.

\subsubsection{Object Identification}
\label{sec:photometry}

Following all of the preprocessing, photometric analysis was carried out using
the {\it hstphot} task, a program that automatically accounts for WFPC2
point-spread function (PSF) variations (using a library PSF calculated from
Tiny Tim [\citealt{Krist95}] models with an added residual calculated
empirically from the data), charge-transfer effects across the chips,
zeropoints, and aperture corrections.  When possible, {\it hstphot} returns
magnitudes in standard Johnson-Cousins photometric bands
\citep{Johnson66,Cousins81} as output.  We ran {\it hstphot} with option flag
10, which combines turning on local sky determination and turning
off empirically determined aperture corrections.  Turning on local sky
determination is recommended by the HSTphot manual for images with rapidly
varying backgrounds, while turning off aperture corrections is necessary since
the NGC~1637 field does not contain enough bright and isolated stars to
determine a reliable empirical correction.  By turning off empirical aperture
corrections, default values for each filter are applied to the photometry, and
are accurate, in general, to 0.02 mag (see \S~\ref{sec:cvatplr}).  We ran {\it
hstphot} with a $1\ \sigma$ detection threshold for both the combined and
individual epochs, opting for such a low signal-to-noise (S/N) threshold in
order to mitigate potential selection bias against fainter objects.  With these
input parameters, {\it hstphot} identified 74,307 objects in the NGC~1637
field.

\subsubsection{Comparing HSTphot with other Photometry Packages}
\label{sec:cbhaopp}

Photometry produced by HSTphot has been compared with that generated by other
packages, including DoPHOT \citep*{Schechter93} and DAOPHOT/ALLFRAME
\citep{Stetson87,Stetson94}, and the results show excellent agreement
\citep{Dolphin00a,Saha01a}.  This is not surprising, since most of the
machinery of HSTphot, including the star-finding and PSF-fitting algorithms,
are in fact modeled after these older packages. The primary advantage of
HSTphot is that it has built-in knowledge of WFPC2 instrumental characteristics
and hence runs with far less user interaction and produces robust results that
are easily reproducible by different users.

As a ``sanity check'' on our own implementation of HSTphot, as well as our
technique for identifying and characterizing Cepheid variables
(\S~\ref{sec:results}), we reduced and analyzed the archival {\it HST} WFPC2
data acquired for NGC~3351 as part of the KP's Cepheid study of the galaxy
\citep[][{\it HST} observing program GO-5397]{Graham97}, using {\it exactly}
the same techniques that were employed for the NGC~1637 data.  One comparison
of the photometry can be made with the 25 bright, nonvariable, comparison stars
listed in Table~3 of \citet{Graham97}.  The photometry reported by the KP was
produced by DAOPHOT/ALLFRAME.  In the $V$ and $I$ passbands, we derive mean
differences of
\begin{displaymath}
 \begin{array}{lll}
  <\Delta V> & = & 0.015 \pm 0.022 {\rm\ mag}, \\
  <\Delta I> & = & 0.036 \pm 0.018 {\rm\ mag},\\
 \end{array}
\end{displaymath}
\noindent where $\Delta V \equiv V_{\rm HSTphot} - V_{\rm KP}$, $\Delta I
\equiv I_{\rm HSTphot} - I_{\rm KP}$, and the error is the standard uncertainty
in the mean.  While the agreement is encouraging, such a comparison is somewhat
obfuscated by many subtle changes in the handling of CTE effects, aperture
corrections, PSF fitting, zero-points, and so forth, that are incorporated in
the recent version of HSTphot that were not available when the original KP
measurements were made.  More directly comparable are the final,
metallicity-corrected Cepheid distances to NGC~3351: $\mu_0 {\rm\ (HSTphot)} =
30.13 \pm 0.07$ (statistical) mag, and $\mu_0 {\rm\ (KP)} = 30.00 \pm 0.09$ (statistical)
mag (F01).  The favorable comparison of these distance moduli provides some
assurance that our photometry and analysis techniques, although somewhat
different in detail from those employed by the KP, ultimately yield similar
results.

\section{RESULTS}
\label{sec:results}

\subsection{Photometry of Comparison Stars}
\label{sec:results1}
Following in the tradition established by the KP, we think it wise to list
measurements of a sample of bright, unsaturated, non-variable stars on all four
WFPC2 chips that can be easily remeasured at a later date for comparison.  In
Table~\ref{tab:tab4}, we list the positions and magnitudes of several such
reference stars for each chip, along with the object number assigned to each
star by {\it hstphot}.  The reported positions are those of the final,
combined, $V$-band frame; the positions of stars on the four WFPC2 chips are
extremely consistent throughout the entire series of observations, with the
maximum linear shift between any two epochs amounting to only $\sim 1/2$ of a
pixel.  (Note that the X and Y positions reported by {\it hstphot} are made so
that an integer value is assigned to a star that is centered in the lower left
corner of a pixel; this is the same convention as DoPHOT, but 0.5 lower in both
X and Y than DAOPHOT.)

\subsection{The Late-Time Lightcurve of SN 1999em}
\label{sec:results2}

A wonderful benefit of obtaining Cycle 10 {\it HST} observations of NGC~1637 is
that it permits a study of the late-time photometric behavior of SN~1999em
itself.  At $\sim 2$ years after the explosion, these are the latest
photometric observations ever obtained for a normal SN~II-P.  During the
nebular phase, an SN~II-P is expected to have an optical lightcurve powered
primarily by thermalized gamma rays resulting from the radioactive decay of
$^{56}{\rm Co}$ into $^{56}{\rm Fe}$ (\citealt{Woosley88}; $^{56}{\rm Co}$
itself is the daughter of $^{56}{\rm Ni}$, which has a half-life of only 6.1
days.)  An analysis of earlier nebular-phase photometry and spectroscopy of
SN~1999em is given by \citet{Elmhamdi03}, who demonstrate that SN~1999em is
indeed fading in optical brightness at a rate consistent with the decay slope
of $^{56}{\rm Co} \rightarrow$ $^{56}{\rm Fe}$ of $0.98\ {\rm mag\ (100\
d)}^{-1}$ during the time covered by their observations, which sample $ 180\
{\rm to}\ 508$ days after explosion.  A linear least squares fit to the
photometric observations of \citeauthor{Elmhamdi03} yields slopes of $\sim
0.97\ {\rm\ mag\ (100\ d)}^{-1}$ and $\sim 1.07 {\rm\ mag\ (100\ d)}^{-1}$ for
the decline in $V$ and $I$, respectively.  In the spectral data,
\citeauthor{Elmhamdi03} note an abrupt blueshift of spectral line profiles
between days 465 and 510, and interpret this as resulting from the formation of
dust in the SN atmosphere; an associated drop in optical brightness is also
seen at this time.  Similar spectroscopic and photometric behavior was observed
in SN~1987A near 500 days after explosion as well, and was given a similar
explanation \citep{Danziger89}.  Between days 500 and 750, SN~1987A exhibited
an accelerating rate of decline of optical brightness along with an
accompanying increase in brightness at infrared wavelengths.  The spectroscopic
and photometric behavior of SN~1987A during this later time have been
attributed to the combined effects of newly formed dust, which extinguishes
optical light and contributes thermal infrared radiation, and an increasingly
transparent (to gamma rays) envelope \citep[see][and references
therein]{Suntzeff92}, which allows radioactive decay energy to escape without
being thermalized.  The dust is postulated to exist in many small, optically
thick clumps, which produces nearly gray extinction \citep{Lucy88}.

Our {\it HST} observations provide 12 $V$ and 6 $I$ epochs sampling $679\ {\rm
to}\ 738$ days after the date of explosion of SN~1999em estimated by
\citet{Leonard5}.  The results are listed in Table~\ref{tab:tab5}.  A
comparison between the absolute $V$ and $I$ magnitudes of SN~1999em and
SN~1987A is shown in Figure~\ref{fig:plot13}, where the observed SN~1987A data
of \citet{Hamuy90} have been corrected for $\Ebv = 0.15$ mag \citep{Arnett89}
and put on an absolute scale by assuming $\mu_{\rm LMC} = 18.50$ mag
(\S~\ref{sec:cvatplr}); the SN~1999em data have been corrected for $\Ebv =
0.10$ mag \citep{Leonard5} and put on an absolute scale by using the distance
modulus to NGC~1637 derived in \S~\ref{sec:tcdtns} (note that this distance is
also ultimately based on the assumption that $\mu_{\rm LMC} = 18.50$ mag;
\S~\ref{sec:cvatplr}).  Similar to SN~1987A at this phase, the decline rates in
$V$ and $I$ significantly exceed the $^{56}{\rm Co}$ decay slope: a
least-squares fit to our data yields decline slopes of $1.62 \pm 0.05 {\rm\
mag\ (100\ d)}^{-1}$ in $V$-band and $2.28 \pm 0.11 {\rm\ mag\ (100\ d)}^{-1}$
in $I$-band.  As was inferred for SN~1987A, this is likely due to increased
envelope transparency to gamma rays, and perhaps also to dust formation in the
cooling SN atmosphere.  Curiously, SN~1999em appears to be getting somewhat
bluer with time during these epochs, within the errors.  Such behavior is not
seen in SN~1987A at these epochs, although it is seen at other times
\citep[see, e.g.,][]{Bouchet96}.  The reason for this behavior is unclear,
although variations in the strength of nebular emission lines may play a role.

The changes in slope of the ``exponential tail'' $V$-band light curve of
SN~1999em appear to have closely followed those of SN~1987A, since it declined
according to the exponential decay slope of $^{56}{\rm Co}$ up until about day
500, and then faded at an increasing rate similar to SN~1987A thereafter.  This
suggests a similar evolution of the fraction of radioactive decay energy
deposited into the envelope (e.g., nearly $100\%$ up through day 500, followed
by increasing transparency thereafter).  At the time covered by our {\it HST}
observations, SN~1999em is $\sim 0.25$ mag brighter than SN~1987A.  Assuming
similar bolometric corrections (e.g., \citealt{Hamuy01a, Hamuy03}) at this
nebular phase for the two events (probably a reasonable assumption since the
$V$ band is essentially a continuum filter at this stage, as there are no
strong emission lines in this spectral region), this implies that $L_{\rm
99em}/L_{\rm 87A} \approx 1.26$.  From detailed modeling, \citet{Arnett96}
estimates the mass of radioactive $^{56}{\rm Ni}$ ejected by SN~1987A to be
$0.075\ M_\odot$.  Since nebular-phase luminosity scales in direct proportion
to the amount of synthesized $^{56}{\rm Ni}$, and SN~1999em and SN~1987A appear
to have had similar energy-deposition rates during the exponential tail, this
implies that $\sim 0.09\ M_\odot$ of $^{56}{\rm Ni}$ was synthesized and
ejected by SN~1999em \citep[cf.][]{Elmhamdi03}.

\subsection{Identification of Variable Stars}
\label{sec:iocvs}

To identify potential variable stars among the 74,307 objects, we tested each
object against a sequence of increasingly stringent selection criteria.  The
choice of selection criteria was guided by the philosophy that it is better to
lose some real Cepheids from the sample than to include spurious or doubtful
objects.  We were therefore very conservative in the selection criteria used to
produce the final Cepheid sample.  The selection criteria, in the order that
they were applied, are as follows.

1. Object must be classified as a ``point source'' (as opposed to, e.g., an
   extended object or unresolved binary) by {\it hstphot}.  This step
   eliminated 3,453 objects from further consideration.

2. Object must be more than a PSF radius away from the edge of the unvignetted
   portion of each chip.  This eliminated 831 objects.

3. Object must be well fit by the PSF-fitting routine within {\it hstphot},
   with reported values of $\chi < 1.35$ and $\mid{\rm sharpness} \mid <
   0.4$. This eliminated 11,327 objects.

4. Object must have photometry reported for at least $9\ V$ and $3\ I$ epochs.
   This eliminated 22,739 objects.

5. Object must have a $99.9\%$ chance of being variable based on the photometry
   and photometric uncertainty reported for all of the photometric epochs for
   the F555W filter ($V$-band).  For this test, we applied the criterion of
   \citet{Saha90} that $P(\chi^2_\nu) < 0.001$, where
\begin{equation}
P(\chi^2_\nu) \equiv  \int_{\chi^2_\nu}^{+\infty} \frac{x^{2[(\nu/2) -1]} e^{-x^2/2}}{2^{\nu/2}\Gamma(\nu/2)} dx^2 ,
\end{equation}
\noindent and $\nu$ is the number of degrees of freedom ($\nu = n - 1$, where
$n$ is the number of epochs with reported photometry), $\Gamma$ is the gamma
function, and
\begin{equation}
\chi^2 = \frac{1}{n-1}\sum_{i=1}^n \frac{(m_i - \overline{m})^2}{\sigma_i^2},
\end{equation}
\noindent where $m_i$ is the magnitude of the $i$th exposure with associated
uncertainty $\sigma_i$, and $\overline{m}$ denotes the average over the $n$ values
of $m_i$.  This criterion eliminated 32,221 objects from further
consideration, leaving 3,736 candidate variable stars.

\subsection{Identification of Cepheid Variables}
\label{sec:ioccv}

To distinguish Cepheids from other variables, we next fit (using weighted
least-squares fitting, with weights based on the {\it hstphot} reported errors)
each candidate variable with empirical template light curves based on a
principal component description of the light curves of nearby Cepheids
\citep[see, e.g., ][]{Kanbur02}.  The training data set consists of $V$- and
$I$-band photometry of about 50 Cepheids with $10 < P < 80$ days in the
Magellanic Clouds and Milky Way.  The photometry comes mainly from
\citet{Moffett84}, \citet{Berdnikov95b}, \citet{Moffett98}, and
\citet{Tanvir99b}, and is well sampled with small photometric errors.  These
light curves (both bands) were Fourier analysed to 12 terms \cite[e.g.,
][]{Ngeow03}, and the principle component analysis applied to vectors
consisting of all of the Fourier coefficients from each band.  The resulting
principal component coefficients are found to be functions of period, and thus
at any period we can use them to determine a range of typical light curves.

To find the period with the best-fitting template, we fit the $V$ and $I$ data
points simultaneously, searching periods from 10 to 60 days.  Setting the lower
period bound at 10 days avoids contamination by first overtone pulsators and
extrapolation of the PCA coefficients beyond the range defined by the template
sample.  The upper period limit of 60 days is set by the baseline of our
observations.  To arrive at the best-fitting template for each period, we
solved for the best values of five free parameters: (1) phase offset between
the first data epoch and the template light curve; (2) mean $V$ magnitude; (3)
mean $I$ magnitude; (4) magnitude of the first PCA coefficient, constrained to
fall within $1.5\ \sigma$ of the ridgeline established by the training-set
Cepheids; and (5) magnitude of the second PCA coefficient, constrained to fall
within $1.5\ \sigma$ of the ridgeline established by the training-set Cepheids
(see also \citealt{Ngeow03} for the Fourier analogue of this technique).  Upon
arriving at the period with the best-fitting template for each candidate
variable, we then tested and eliminated objects from the sample based on the
following additional selection criteria (continuing the numbering from
\S~\ref{sec:iocvs}):

6. Best-fitting template must have a ``relative likelihood'' $> 0.3$, where
``relative likelihood'' is defined as the quantity $e^{-\chi^2/2\nu}$, with
$\nu$ representing the number of degrees of freedom of the fit (e.g., $\nu =
12$ for a candidate with data for all 12 $V$ and 6 $I$ epochs, since
there are six lost degrees of freedom from the fitting procedure), and 
\begin{equation}
\chi^2 \equiv \sum_{i=1}^n \frac{(m_i - m_{\rm template})^2}{\sigma_i^2},
\end{equation}
\noindent
where $m_i {\rm\ and\ } \sigma_i$ are the data points and uncertainty, respectively,
 at each epoch $i$, and $m_{\rm template}$ is the value of the best-fitting template
 lightcurve corresponding to data epoch $i$.  Objects with best-fitting
 templates that fail this criterion are considered to not have appropriately
 ``Cepheid-like'' light curves.  This test eliminated 1415 objects.

7. Object must possess:
\begin{equation}
\frac{ {\rm Highest\ relative\ likelihood} } { {\rm Median\ relative\ likelihood} }
> 50,
\label{eqn:twenty}
\end{equation}
\noindent 
over the tested period range of $10 - 60$ days.  This test was very effective
at eliminating objects with period aliases, ultimately removing 2,279 objects
from further consideration.  However, while period aliasing may result from
incomplete phase coverage of even bright, well-measured objects, it is
important to note that aliasing is also a natural consequence of low S/N
measurements, for which (appropriately) large error bars produce a wide range
of acceptable periods.  For our WFPC2 photometry of NGC~1637, this begins to
occur near $V \approx 26$ mag, at which point the standard error of a magnitude
determination becomes comparable to the RMS variation in the lightcurve of a
Cepheid.  Therefore, while this test undoubtedly eliminates many non-Cepheids,
it also contains a selection bias against fainter, but potentially real,
Cepheids.  Such a selection bias has the potential to yield a final sample in
which bright objects are overrepresented, especially at the shorter (and,
hence, fainter) periods.  Such an ``incompleteness bias'' at the faint end of a
target galaxy's apparent PL relations has been seen in previous {\it HST}
Cepheid studies (see, e.g., F01; \citealt{Saha01b}), and we must be watchful
for its presence in our final Cepheid sample.  The effect of this bias depends
in a complex way on the selection criteria and level of correlated noise
between the bands.  Fortunately, the incompleteness bias usually reveals itself
empirically, and is rather easily remedied by establishing a lower period
cutoff for inclusion in the final set of Cepheids that are used to determine
the distance to the target galaxy.  The cutoff period is determined by finding
the point at which the apparent distance moduli (eq.~[\ref{eqn:two}]) in both
bands stabilize.

8. Object must have no significant (relative likelihood exceeding $0.3$)
aliases outside of the target range of $10 < P < 60$ days.  To check for
potential short- or long-period aliases, we searched the remaining sample of 42
objects for period aliases over the extended range of 5 to 100 days.  One
object was eliminated from further consideration by this test.

Our selection process has therefore yielded 41 viable Cepheid candidates with
well-determined periods in the range of $10 < P < 60$ days.  They are
identified in Table~\ref{tab:tab6}.  Finding charts are given in
Figures~\ref{fig:2} and \ref{fig:3}, and light curves are shown in
Figure~\ref{fig:4}.  Complete photometry of the variables is given in
Table~\ref{tab:tab7}.

As discussed in \S~\ref{sec:cvatplr}, to determine the apparent $V$ and $I$
distance moduli, $\mu_V$ and $\mu_I$ (eq.~[\ref{eqn:two}]), as well as the
unreddened distance moduli, $\mu_W$ (eq.~[\ref{eqn:eight}]), we adopt the PL
relations of F01:
\begin{equation}
 \begin{array}{lll}
  M_V  & = & -2.760 \log(P) - 1.458,\\
  M_I  & = & -2.962 \log(P) - 1.942,\\
 \end{array}
\label{eqn:twentyone}
\end{equation}
\noindent which, through equation~(\ref{eqn:nine}), yield
\begin{equation}
M_W = -3.255 \log(P) - 2.644.\\
\label{eqn:twentytwo}
\end{equation}

\noindent As discussed in \S~\ref{sec:cvatplr}, these PL relations were derived
by F01 from photometry of $\sim 650$ LMC Cepheids that were observed as part of
the OGLE project \citep{Udalski99}, which represents the largest study of LMC
Cepheids.  Given $\mu_V$ and $\mu_I$ for each Cepheid, it is also possible to
solve for the extinction through the relation
\begin{equation}
A_V = R(\mu_V - \mu_I),
\label{eqn:twentythree}
\end{equation}
\noindent where $R = 2.45$ (see discussion in \S~\ref{sec:cvatplr}).  In
Table~\ref{tab:tab8}, we list the periods, mean $V$ and $I$ magnitudes
(determined as intensity-mean magnitudes, from direct integration of the
best-fitting template light curves), $A_V$, $\mu_V$, $\mu_I$, and $\mu_W$ for
the Cepheid candidates.  A color-magnitude diagram of all of the stars in the
NGC~1637 field is shown in Figure~\ref{fig:6}.  The Cepheids populate the
instability strip, with some scatter likely due to reddening or measurement
errors.

\subsection{The Cepheid Distance to NGC 1637}
\label{sec:tcdtns}

From the 41 Cepheids, we derive $\overline{\mu_V} = 30.60 \pm 0.05$
(statistical) mag and $\overline{\mu_I} = 30.39 \pm 0.04$ (statistical) mag,
which implies an average extinction of $\overline{A_V} = 0.49$ mag through
equation~(\ref{eqn:twentythree}).  This then yields a reddening-corrected
distance modulus of $\overline{\mu_W} = 30.10 \pm 0.05$ (statistical) mag from
equation~(\ref{eqn:four}).

The apparent PL relations in $V$ and $I$ for the 41 Cepheid variables are shown
in Figure~\ref{fig:7}(a, b).  Examination of these figures suggests that our
Cepheid sample may be affected by the incompleteness bias discussed in
\S~\ref{sec:ioccv}, since there is a preponderance of data points above the
ridge lines at short periods in both bands.  To test for incompleteness bias we
examined how the derived apparent distance moduli for $V$ and $I$ change as a
function of the lower cutoff period.  The results are shown in
Figure~\ref{fig:8}(a, b).  The steady rise in the apparent distance moduli for
cutoff periods below $23.0$ days suggests that we are affected by
incompleteness bias at these shorter periods.  The degree of any bias becomes
very small for cutoff periods larger than $23.0$ days.  We therefore adopt a
lower period cutoff of $P_{\rm cut} = 23.0$ days ($\log[P_{\rm cut}] = 1.36$
days) for inclusion in the final sample, which leaves 18 Cepheids from which to
derive the distance to NGC~1637.

The reddening-free PL relation is shown in Figure~\ref{fig:7}c.  Taking the
unweighted mean of the unreddened distance moduli for the 18 Cepheids with $23
< P < 60$ days yields an unreddened distance modulus of $\overline{\mu_W} =
30.23 \pm 0.07 \pm 0.17 {\rm\ mag}$, where the systematic uncertainty is the
value derived in \S~\ref{sec:cvatplr} and shown in Table~\ref{tab:tab2}.  Our
correction for incompleteness bias, which amounts to $0.13$ mag ($\sim 6\%$ in
distance; see Fig.~\ref{fig:8}c) is somewhat larger than those typically derived
in {\it HST} Cepheid studies (see, e.g., F01).  We suspect that this is 
due to the conservative cut applied to the data by selection criterion 7
(\S~\ref{sec:ioccv}), which removes objects with significant period aliases
from the sample.  As previously discussed, this criterion begins to select
against fainter objects at the point where the error of a magnitude
determination begins to be comparable to the RMS variation in the light curve
of a Cepheid.  One way to mitigate this unwanted effect is to lower the
threshold for acceptance defined by equation~(\ref{eqn:twenty}).  However,
while lowering this threshold does result in a larger number of Cepheid
candidates being included in the final sample, the sample becomes compromised
by dubious objects that sometimes yield very discrepant distances.  Rather than
resorting to subjective criteria or applying arbitrary cuts based on derived
distance or color that would, almost inevitably, still permit some non-Cepheids
to dilute the final sample, we feel that the conservative, but quantitative and
objective, approach is best.  In this effort we are aided by the rather large
sample of Cepheids identified in the target galaxy, which enables a
statistically sound sample of {\it bona fide} Cepheids to remain after the
lower period cutoff is applied.

Applying the metallicity correction of $\delta\mu_Z = 0.12 \pm 0.12$ mag
derived in \S~\ref{sec:cvatplr} yields a final Cepheid distance to NGC~1637 of:
\begin{eqnarray*}
  \mu_0 & = & 30.34 \pm 0.07 \pm 0.17 {\rm\ mag,\ or}\\
   D    & = & 11.71  \pm  0.36  \pm  0.92  {\rm\ Mpc,}
\end{eqnarray*}
through equation~(\ref{eqn:twelve}). It is immediately apparent that this
distance is significantly discrepant with the EPM distance derived earlier
(\S~\ref{sec:tepmdtsnn}), in the sense that {\it the Cepheid distance to
NGC~1637 is nearly $50\%$ greater than the EPM distance to SN~1999em.}

\section{DISCUSSION}
\label{sec:discussion}

\subsection{Comparison of the Cepheid and EPM Distances to NGC 1637}
\label{sec:discussion1}

The primary goal of this Cepheid study is to test and, ultimately, to calibrate
the SN~II-P distance scale.  To derive our distance to NGC~1637, we have
largely followed the procedures of the KP, as outlined by F01.  As discussed in
\S~\ref{sec:cvatplr}, however, there is currently some debate between the KP
and STS groups on the appropriate values to use for the slopes of the $V$- and
$I$-band PL relations.  Therefore, before concluding that a significant
discrepancy exists between the Cepheid and EPM distance scales, it seems
prudent to check our Cepheid distance against the distance that would have been
derived had we adopted the current methodology of the STS group, as given by
\citet{Thim03} and described in \S~\ref{sec:cvatplr} (e.g., taking the mean of
the distance moduli resulting from the PL relations derived from Cepheid Sets
{\it B, C,} and {\it D} in Table~1).  When this is done, a Cepheid distance of
$D_{\it STS} = 11.81$ Mpc ($\mu_0 {\rm [STS]} = 30.36$ mag) results.  This
distance is in excellent agreement with the KP-derived value, $D_{\it KP} =
11.71$ Mpc, and builds confidence in the robustness of the Cepheid/EPM
discrepancy that we have found (as well as the uncertainty ascribed to the
slope of the PL relations in \S~\ref{sec:cvatplr}).

We therefore conclude that the EPM distance to SN 1999em of $D_{\rm EPM} = 7.86
\pm 0.50$ (statistical) Mpc (\S~\ref{sec:tepmdtsnn}) is indeed inconsistent, at
the $3.5\ \sigma$ level, with the Cepheid distance to NGC~1637 of $D = 11.71
\pm 0.99 {\rm\ Mpc}$.  Since the total uncertainty in the Cepheid distance is
estimated to be $\lesssim 10\%$ whereas the EPM distance may suffer from as
much as a $60\%$ systematic uncertainty ({\S~\ref{sec:tdf}), suspicion
immediately falls on the EPM as having underestimated the distance to
SN~1999em.  This contention is supported by other considerations as well, which
we now discuss.

\subsection{Additional Distance Estimates to NGC 1637}
\label{sec:discussion3}

Seven other estimates of the distance of NGC~1637 exist, six of which favor the
longer Cepheid distance over the shorter EPM distance.

(1) {\it The Tully-Fisher relation.}  Due to its nearly face-on orientation and
   somewhat lopsided disk \citep{Ryder93,Block94}, NGC~1637
   is not an ideal Tully-Fisher (TF) target.  Nonetheless, a $B$-band TF
   distance has been derived by \citet{Bottinelli85}:  
\begin{eqnarray*}
   \mu_0 & = & 30.70  \pm  0.38 {\rm\ mag,\ or}\\
  D   & = & 13.8     \pm  2.4 {\rm\ Mpc.}
\end{eqnarray*}

(2) {\it Inner ring structure.}  By examining the diameter of the inner ring
   structure of NGC~1637, \citet{Buta83} derive a distance to NGC~1637 of
\begin{eqnarray*}
   \mu_0 & = & 30.23 \pm 0.69 {\rm\ mag,\ or}\\
   D         & = & 11.12 \pm 3.53 {\rm\ Mpc.} 
\end{eqnarray*}
\noindent

(3) {\it The kinematic distance.}  Adopting a Hubble constant of $H_0 = 68\ 
\kmsmpc$, and applying the parametric model for peculiar flows of
\citet{Tonry00}, \citet{Hamuy03} derives a kinematic distance to NGC~1637 of
\begin{eqnarray*}
   \mu_0  & = & 30.15  \pm  0.49 {\rm\ (statistical)} {\rm\ mag,\ or}\\
   D   & = & 10.7 \pm 2.4 {\rm\ (statistical)} {\rm\ Mpc.}
\end{eqnarray*}

(4) {\it The spectral-fitting expanding atmosphere method applied to
SN~1999em.}  The dilution factors of E96 (as modified slightly by
\citealt{Hamuy01b}) were used in all three EPM estimates of the distance to
SN~1999em.  As discussed in \S~\ref{sec:tdf}, however, a nagging source of
uncertainty in the EPM technique is the possible disagreement among various
modeling groups of the magnitude of the theoretically derived dilution factors,
which have a profound impact on EPM distances (e.g., eq.~[\ref{eqn:fifteen}]).
In the only direct comparisons that have been made to date, the radiative
transfer models of \citet{Baron94} have consistently yielded dilution factors
that are significantly larger than those of E96, in some cases by as much as
$60\%$ \citep{Mitchell02,Baron95}.  However, the fact that these dilution
factors were produced for somewhat peculiar SNe~II, and not normal SNe~II-P,
casts doubt on the significance of the offset, which will remain until a full,
independent set of dilution factors appropriate for an SN~II-P is produced.

Thankfully, such a project is now underway, and a distance to SN~1999em itself
is currently being derived by E. Baron and collaborators using the PHOENIX
radiative transfer code \citep{Baron94}.  The technique being used to determine
the distance is formally known as the spectral-fitting expanding atmosphere
method (SEAM; \citealt{Baron93}; \citealt{Baron94}; \citealt{Baron95}),
although it is similar in spirit to EPM.  Unlike the standard implementation of
the EPM for which a generic set of average dilution factors for SNe~II-P are
used to correct distances via equation~(\ref{eqn:fifteen}), however, the SEAM
establishes distances through the detailed modeling of individual spectra.  The
fitted models provide the entire spectral energy distribution of an SN on an
absolute scale, from which synthetic absolute photometry is derived.  The
synthetic photometry is then compared with the observed photometry and a
distance modulus for each spectral epoch in each available photometric band
results.  There is no explicit need for dilution factors in the technique,
although they may be calculated for comparison purposes.  A preliminary SEAM
analysis of SN~1999em (E. Baron, private communication), based on spectral fits
to only the two early time spectra analyzed in the study by \citet{Baron00},
yields:
\begin{eqnarray*}
   \mu_0 & = & 31.0 \pm 0.6 \pm 0.5 {\rm\ mag,\ or}\\
   D         & = & 15.8 \pm 4.4 \pm 3.6 {\rm\ Mpc.}
\end{eqnarray*}
The large uncertainty in this preliminary distance will undoubtedly be reduced
as additional spectra are analyzed.  Nonetheless, at this stage, a significant
preference for the longer, Cepheid-based, distance exists, providing some
support for the contention that much of the Cepheid/EPM distance discrepancy
may be explained by an underestimate of the E96 dilution factors for at least
this SN~II-P.

(5) {\it Standard-candle method for SNe~II-P.}  SNe~II-P are {\it not} standard
candles, since they display a range in $V$-band plateau brightness of more than
$5$ mag.  However, a promising empirical technique has emerged that has the
potential to achieve distance estimates of SNe~II-P with a precision rivaling
those currently derived to SNe~Ia (e.g., $\sim 10\%$).  Developed by
\citet{Hamuy02}, the method treats SNe~II-P as ``calibratable standard
candles'' by adjusting individual luminosity differences through the use of an
empirically derived relationship between expansion velocity (as determined from
the blueshift of spectral lines) and luminosity (in the sense that brighter
SNe~II-P have higher expansion velocities).  This type of empirical correction
is quite similar in spirit to the accepted $\Delta m_{15}$ \citep{Phillips93}
and multicolor light-curve shape \citep{Riess96} methods that are used to
adjust SNe~Ia luminosities to a ``standard'' value.  The initial results of the
SCM for SNe~II-P are impressive: from a sample of 16 SNe~II-P, \citet{Hamuy02}
find that the correlation between expansion velocity and luminosity reduces the
scatter in the Hubble diagram from $\sim 1$ mag to $\sim 0.4$ and $\sim 0.3$
mag in the $V$ and $I$ bands, respectively.  Further restricting the sample to
objects that are well in the Hubble flow yields an even tighter result: for
eight SNe~II-P with $cz > 3000\ \Kms$, the Hubble diagram scatter drops to only
$0.2$ mag in both bands, or roughly $9\%$ in distance.  Compared with the EPM,
which has an internal precision of $\sim 20\%$ (\S~\ref{sec:ssu}), the SCM
appears to be significantly more precise.  It is also far easier to apply since
it requires only a few photometric epochs, a single spectrum, and no
theoretical modeling.

Since SCM is not a primary distance indicator, however, the zero-point of its
Hubble diagram must be determined externally.  Of course, prior to our present
study, no host galaxy of a normal, well-observed SN~II-P had its distance
measured by a reliable primary distance indicator (i.e., Cepheid variables).
Therefore, \citet{Hamuy02} cautiously establish the SCM zero point by using
SN~1987A in the LMC (assumed distance modulus of $18.50$ mag), even though
SN~1987A is an admittedly rather peculiar variant of the plateau subclass.
Nonetheless, using the scale established by SN~1987A and the $V$ and $I$
plateau magnitudes and expansion velocities of SN~1999em given by
\citet{Hamuy02},\footnote{Note that the expansion velocity of SN~1999em given
by \citet{Hamuy02}, of $v = 3557\ \Kms$, has a typographical error (M. Hamuy,
personal communication).  The correct value, $v = 3757\ \Kms$, is given by
\citet{Hamuy03}.} yields an SCM distance to SN~1999em of:
\begin{eqnarray*}
   \mu_0  & = & 30.49  \pm  0.40 \pm 0.57 {\rm\ mag,\ or}\\
   D   & = & 12.53 \pm 2.31 \pm 3.29 {\rm\ Mpc,}
\end{eqnarray*}
\noindent from the $V$-band relation, and
\begin{eqnarray*}
   \mu_0  & = & 30.63  \pm  0.29 \pm 0.45 {\rm\ mag,\ or}\\
   D   & = & 13.37 \pm 1.79 \pm 2.77 {\rm\ Mpc,}
\end{eqnarray*}
\noindent from the $I$-band relation.  The quoted systematic uncertainties
incorporate uncertainties in the slope of the calibrating SCM relations (see
eqs.~[1] and [2] of \citealt{Hamuy02}), the extinction ($\sigma [A_V] = 0.3$
mag; $\sigma [A_I] = 0.18$ mag), plateau magnitude ($\sigma [V_p] = 0.05$ mag;
$\sigma [I_p] = 0.05$ mag), photospheric velocity ($\sigma [v_p] = 300\ \Kms$),
and distance modulus of SN~1987A ($\sigma[\mu_0] = 0.1$ mag).  It does not
include the (unknown) systematic uncertainty added by the fact that SN~1987A
was not a typical SN~II-P.  The statistical errors result from the uncertainty
in the extinction ($\sigma [A_V] = 0.3$ mag; $\sigma [A_I] = 0.18$ mag),
plateau magnitude ($\sigma [V_p] = 0.05$ mag; $\sigma [I_p] = 0.05$ mag), and
photospheric velocity ($\sigma [v_p] = 300\ \Kms$) of SN~1999em.  The $V$- and
$I$-band distances yield an average estimate of about $13 \pm 4$ Mpc.  It must
be noted that since the zero points of both the SCM and the Cepheid PL relation
are set by the LMC distance modulus, the ultimate calibrations of the two
techniques are not completely independent, and share the same systematic
uncertainty ($0.1$ mag) contributed by the uncertainty in the LMC distance.

(6) {\it Plateau-tail method for SNe II-P}.  Motivated primarily by theoretical
    considerations, \citet{Nadyozhin03} has recently introduced a novel way to
    estimate distances to SNe~II-P that involves both the plateau and
    exponential tail luminosities as well as the photospheric velocity during
    the plateau.  For SN~1999em, he finds
\begin{eqnarray*}
   \mu_0  & = & 30.2  \pm  0.4 {\rm\ (statistical)\ mag,\ or}\\
   D   & = & 11.1 \pm 2.2 {\rm\ (statistical)\ Mpc,}
\end{eqnarray*}
\noindent where the uncertainty has been estimated from the average scatter of
the plateau-tail distances derived by \citet{Nadyozhin03} about the Hubble line
for the eight SNe~II-P considered in the study.  These SNe~II-P also have
earlier EPM distance estimates, and it is interesting to note that
\citet{Nadyozhin03} finds an average distance discrepancy between the two
techniques of $\sim 25\%$, in the sense that $D_{\rm EPM}$ is systematically
{\it shorter} than $D_{\rm plateau-tail}$.

(7) {\it Brightest red supergiants.}  Treating the brightest red
   supergiants (BRSG) detected in ground-based images of NGC~1637 as
   standard candles, \citet{Sohn98} derive
\begin{eqnarray*}
   \mu_0  & = & 29.47  \pm  0.27 {\rm\ (statistical)\ mag,\ or}\\
   D   & = & 7.8^{+1.0}_{-0.9} {\rm\ (statistical)\ Mpc,}
\end{eqnarray*}
\noindent where the stated error accounts for the difficulty of
identifying the BRSG and a correction for crowding effects.  

Since the BRSG technique provides the only distance estimate that favors the
shorter EPM distance to NGC~1637 over the longer one determined by Cepheids, it
is important to quantify its potential systematic uncertainty.  The merit of
the BRSG technique is that red supergiants are among the very brightest stars
in a galaxy, and can be detected in a single epoch of observations.  The
drawbacks are that individual red supergiants can easily be confused by
foreground stars, \ion{H}{2} regions, and clusters in the target galaxy, and
they are always few in number leading to weak statistical significance in most
cases.  Not surprisingly, this technique has not been extensively used for
distance determination in recent years.  The basis for the BRSG being standard
candles is primarily empirical, and according to \citet{Rozanski94} systematic
uncertainties for individual galaxy distances are at least $0.5$ mag in
distance modulus.  Adding this in quadrature to the quoted statistical error
suggests that a total uncertainty of at least $0.57$ mag should be adopted for
the \citet{Sohn98} estimate.

Of course, none of the preceding seven distance estimates to NGC~1637 are
individually unassailable.  Taken together, however, the fact that all but one
prefer the Cepheid distance certainly provides support for the longer,
Cepheid-based distance to NGC~1637.  A summary of the distance measurements is
given in Table~\ref{tab:tab9}.

\subsection{The Progenitor Star Mystery}
\label{sec:discussion2}

\citet{Smartt02} examine pre-explosion images of NGC~1637 and use the lack of a
progenitor-star detection in the prediscovery frames, together with an assumed
distance of $D = 7.50$ Mpc from the \citet{Hamuy01b} EPM study, to determine a
detection threshold for the absolute brightness of the progenitor star of
SN~1999em.  \citet{Smartt02} then compare this detection threshold with the
results of detailed stellar evolution calculations based on the most recent
version of the \citet{Eggleton71,Eggleton72,Eggleton73} evolution program,
which traces stellar luminosities through the carbon-burning phase and predicts
final luminosities of core-collapse SN progenitors.  From these models,
\citet{Smartt02} conclude that any progenitor star with $M > 7\ M_\odot$ should
be detectable in the pre-explosion images.  The lack of such a detection at the
location of SN~1999em in these images, especially in light of the theoretical
expectation that core-collapse SNe only result from progenitors with $M \gtrsim
8 - 10 \ M_\odot$ \citep[see][and references therein]{Woosley86}, is quite
puzzling.  Indeed, it is only by considering the {\it uncertainty} on the
threshold detection limit ($\sim 0.2$ dex), that \citet{Smartt02} are able to
(barely) explain the nondetection, and then only for a very narrow range of
possible progenitor masses, $12 \pm 1\ M_\odot$.  However, this calculation
rests squarely on the distance assumed for NGC~1637.  Adopting the longer,
Cepheid-based distance to NGC~1637 derived here will naturally result in a
brighter intrinsic detection threshold for the progenitor star.  In fact,
substituting the Cepheid distance to NGC~1637 of $11.71$ Mpc results in a
$0.39$ dex increase in the detection threshold of the pre-explosion images (see
\citealt{Smartt02}, Figs. 4 and 5).  Adopting the $0.2$ dex uncertainty on the
threshold level and following exactly the same procedure as \citet{Smartt02},
we find that all progenitors with $M < 20\ M_\odot$ could remain undetected in
the pre-explosion images.  Therefore, by adopting the new Cepheid distance to
NGC~1637, we derive an upper mass limit for the progenitor of SN~1999em of $20
\pm 5\ M_\odot$; following \citet{Smartt02}, the uncertainty is simply set by
the nearest modeled progenitor masses on either side of the observed limit.

The details of late-time stellar evolution are still quite controversial, and
subtle changes in the evolution can lead to different conclusions.  For
instance, if we adopt the Geneva evolutionary tracks of \citet{Meynet94},
instead of the models of \citet{Smartt02}, a more restrictive upper limit of $M
< 15^{+5}_{-3}\ M_\odot$ results.  The exact upper limit therefore remains
somewhat uncertain.  Nonetheless, a longer distance to NGC~1637 certainly helps
to explain the absence of a progenitor star in pre-explosion images of the site
of SN~1999em.

\subsection{Potential Causes of the EPM Distance Underestimate}
\label{sec:cftedu}

The preceding discussion provides significant circumstantial evidence to favor
the Cepheid distance over the EPM measurement.  A most important additional
consideration, of course, is that the Cepheid distance scale has already passed
a stringent external check of its accuracy through its comparison with the
maser distance to NGC~4258 (\S~\ref{sec:introduction}), whereas the EPM
distance scale has not.  The case for the shorter EPM distance therefore seems
to us to be difficult to maintain: it agrees with only one of seven other
distance estimates and leaves the puzzling non-detection of the progenitor star
in potential disagreement with the results of current stellar evolution theory.
Furthermore, unlike the case for the EPM, no ready explanation exists for a
major alteration in the Cepheid distance scale, for which systematic
uncertainties are believed to be $\lesssim 10\%$ (\S~\ref{sec:cvatplr}); such
an adjustment to the Cepheid scale, by inference, would also call into question
the quite secure maser distance to NGC~4258.  Therefore, on balance we find
that the available evidence substantially favors the Cepheid distance over the
EPM distance.

The obvious question, then, is why the EPM has underestimated the distance to
SN 1999em.  Since analysis of the scatter in the EPM Hubble diagram leads to
the conclusion that the total statistical uncertainty in a typical EPM
measurement is only $\sim 20\%$ (\S~\ref{sec:ssu}), the likelihood of some
source of statistical error being predominantly responsible for the EPM
underestimate is rather remote.  Of course, it is always possible that
SN~1999em was in some heretofore unsuspected way an unusual SN~II-P, or that
several statistical sources of uncertainty have conspired to produce the
result.  For instance, one scenario that could plausibly explain the distance
underestimate is if SN~1999em was, in fact, highly aspherical during the
plateau phase.  \citet{Nadyozhin98} demonstrates that an SN shaped as an oblate
ellipsoid with an axial ratio of $\sim 0.6$ and oriented such that the symmetry
axis is directed nearly along the line-of-sight (e.g., nearly ``face-on''),
would yield an EPM distance that underestimates the true distance by $\sim
50\%$.  This particular viewing orientation would lead to very little observed
polarization \citep[e.g.,][]{Hoflich91}, and would therefore be consistent with
the spectropolarimetric results of \citet{Leonard3} as well.  The difficulty
with this explanation, however, is that it requires SN~1999em, by all other
accounts a ``normal'' SN~II-P, to be unique: if SNe~II-P as a class are indeed
as aspherical as suggested by this explanation, then random viewing
orientations would yield both larger average intrinsic polarizations and
significantly more scatter in the EPM-based Hubble diagram than are observed.
Since there is at present no compelling indication that SN~1999em was anything
but a ``normal'' SN~II-P, we believe asphericity (or, indeed, any single
statistical source of uncertainty) to be an unlikely sole agent for the EPM
distance underestimate.  The size of the discrepancy between the EPM and
Cepheid distances to NGC~1637 therefore compels us to believe that systematic
error also contributes, with the most likely sources being the dilution factor
and the technique used to estimate (and interpolate) photospheric velocity
(\S~\ref{sec:tdf}).

\subsection{The Value of $H_0$ from SNe~II-P}
\label{sec:tvohftsfst}

The Cepheid distance to SN~1999em allows the calibration of both the EPM and
SCM Hubble diagrams, from which Hubble's constant may be derived.  For the EPM,
the most recent estimate of $H_0$ is that of \citet{Hamuy01a}.  In this study,
nine well-observed SNe~II-P (including SN~1999em) are each assigned EPM
distances that are derived in a consistent manner using the $BV$, $BVI$, and
$VI$ filter combinations to estimate color-temperature and the dilution factors
of E96 (as modified slightly by \citealt{Hamuy01b}). Photospheric velocities
are estimated by two different techniques, and a power-law is fit to the
velocity measurements in order to interpolate them to the epochs of the
photometry (this differs from the technique followed by \citealt{Hamuy01b} in
which a polynomial was fit to the velocities).  The use of three different
filter combinations and two different sets of velocity estimates naturally
results in six different distance estimates for each SN in the study and, hence, six
different Hubble diagrams, from which \citet{Hamuy01a} derives values of $H_0$
ranging from $67\ \kmsmpc$ to $76\ \kmsmpc$; a ``best'' value is not given.
Since at this time we have no reason to favor any particular filter set or
velocity determination method, it seems reasonable to average the six values to
arrive at the estimate given in \S~\ref{sec:introduction}, of $H_0 = 71 \pm 9\
{\rm(statistical)}\ \kmsmpc$.

To derive a new EPM-based value of $H_0$, we therefore calibrate the Hubble
diagrams of \citet{Hamuy01a} using the Cepheid distance to SN~1999em derived
here, and arrive at an average value of 
\begin{eqnarray*}
H_0{\rm (EPM)\ } & = & 57 \pm 7 \pm 13\ \kmsmpc
\end{eqnarray*}
\noindent from the six resulting values of $H_0$.  The statistical uncertainty
in $H_0{\rm (EPM)}$ is derived from the statistical scatter of the SNe~II-P
about the Hubble line.  The systematic uncertainty incorporates the scatter of
the six individual values of $H_0$ and the uncertainty in the offsets between
the Cepheid distance derived in this work and the six EPM distances to
SN~1999em derived by \citet{Hamuy01a}, which includes contributions from the
statistical error in the EPM distances to SN~1999em (for this we adopt the more
conservative $20\%$ uncertainty discussed in \S~\ref{sec:ssu}) and the total
uncertainty in the Cepheid distance to NGC~1637.  It is important to point out
that applying the NGC~1637 calibration essentially reduces EPM to the status of
a secondary distance-determining method, with its ultimate calibration now
provided by Cepheid variables.

For the SCM, calibrating the 16 SNe~II-P in the sample analyzed by
\citet{Hamuy02} with the Cepheid distance to SN~1999em yields $H_0 = 58 \pm 3
\pm 12\ \kmsmpc$ for the $V$-band relation and $H_0 = 60 \pm 3 \pm 10\ \kmsmpc$
for the $I$-band relation.  Taking the average of the two values yields
\begin{eqnarray*}
H_0{\rm (SCM)\ } & = & 59 \pm 3 \pm 11\ \kmsmpc.
\end{eqnarray*}
\noindent The calibrated $V$-band SCM Hubble diagram for this sample of SNe is
shown in Figure~\ref{fig:9}.  The statistical uncertainty in the Hubble
constant derived from SCM is particularly impressive, and rivals that derived
from SNe~Ia (see, e.g., F01).  The SCM therefore appears to be a promising
extragalactic distance-determining method.  However, true confidence in the
technique requires additional, nearby SNe~II-P to be calibrated by the
``training set'' provided by the original set of objects studied by
\citet{Hamuy02}.  Should the calibrating relations prove to be robust, the SCM
could provide precise distance moduli beyond $z \sim 0.3$, and allow an
independent check on the evidence from SNe~Ia for a non-zero cosmological
constant.

The estimates of $H_0$ by both the EPM and SCM would greatly benefit from
additional Cepheid calibrations.  For the EPM, such calibrations will
help to determine whether the Cepheid/EPM discrepancy derived here is truly
universal for SNe~II-P, or whether SN~1999em was in some way a (heretofore
unrecognized) peculiar EPM object.  For the SCM, additional calibrations could
ultimately reduce the uncertainty in $H_0$ to the $\sim 10\%$ level currently
set by the systematic uncertainty in the Cepheid distance scale.  Incidentally,
we note that the Hubble constants derived from the SCM calibration provided by
SN~1987A, $H_0 = 54\ \pm\ 13\ \kmsmpc$ and $H_0 = 53\ \pm\ 10\ \kmsmpc$ for the
$V$ and $I$ relations \citep{Hamuy02}, respectively, are in good agreement with
those derived using our Cepheid NGC~1637 calibration.  This suggests that the
standard formulation of SCM may even be applicable to those SNe~II that show
photometric and spectroscopic peculiarity similar to SN~1987A.

The longer distance scales, and correspondingly smaller Hubble constants,
established by both the EPM and SCM Hubble diagrams when calibrated with the
Cepheid distance to NGC~1637 find good agreement with the conclusions of the
STS group, who report $H_0 = 58 \pm 6\ \kmsmpc$ (T02) from the Cepheid
calibration of SNe~Ia peak luminosity.  The SNe~II-P results are marginally
inconsistent with the results of the KP, who derive $H_0 = 72 \pm 8\ \kmsmpc$
(F01), from the Cepheid calibration of a variety of secondary distance
indicators.  Additional Cepheid calibrations to galaxies hosting SNe~II-P will
allow more statistically stringent comparisons to be made.

\section{CONCLUSIONS}
\label{sec:conclusions}

SN~1999em is an extremely well-studied Type II-P event, for which three
independent and mutually consistent EPM distances have been derived; taking the
average of the three values yields $D_{\rm EPM} = 7.86 \pm 0.50\ {\rm
(statistical)\ Mpc}$.  We present the analysis of multi-epoch WFPC2 Cycle 10
{\it HST} observations of NGC~1637, the host galaxy of SN~1999em, which 
were taken in an effort to derive its distance through the
analysis of its Cepheid variable stars.  This provides the first direct
comparison between the EPM and Cepheid techniques for a galaxy
hosting a well-observed, spectroscopically and photometrically normal, SN~II-P.
As a bonus, the {\it HST} images also contained the fading SN~1999em itself,
roughly two years after it exploded, permitting an analysis of its
nebular-phase photometric behavior.  Our main results are as follows.

1. We identify 41 Cepheid variables in NGC~1637 with $10 < P < 60$ days. To
   avoid the effects of an ``incompleteness bias'' that is suggested by our
   data at short periods, we remove the 23 Cepheids with $P < 23$ days from
   the sample used to determine the final distance to NGC~1637.

2. Using the PL relations of F01, the 18 Cepheids with $23 < P < 60$ days yield
   a mean distance of $11.10 \pm 0.34 \pm 0.87$ Mpc for NGC~1637, before
   correcting for the effects of the metallicity difference between NGC~1637
   and the LMC.  Applying the metallicity correction procedure of F01 yields a
   final, metallicity-corrected, Cepheid distance to NGC~1637 of $11.71 \pm
   0.36 \pm 0.92 {\rm\ Mpc}$.  This distance is nearly $50\%$ greater than the
   previous EPM distances to SN~1999em, and formally represents a $3.5\ \sigma$
   discrepancy between the results.

3. We find compelling evidence to favor the Cepheid distance over the EPM
distance, including the fact that it (a) is in better agreement with six of
seven other distance estimates to NGC~1637; (b) helps to resolve a mystery
concerning the lack of a detectable progenitor star in pre-explosion images of
NGC~1637; (c) is believed to suffer from much lower systematic uncertainty than
the EPM distance ($\sim 10\%$ for Cepheids compared with up to $\sim 60\%$ for
the EPM); and (d) is derived using a technique that has already passed a
stringent external test of its accuracy (e.g., the maser distance to NGC~4258).
We suspect that a substantial portion of the Cepheid/EPM discrepancy may be
explained by an underestimate of the theoretically derived dilution factors of
E96 that have been used in previous EPM distance estimates to normal SNe~II-P,
although other sources of statistical and systematic uncertainty may contribute
as well.

4. Using the Cepheid calibration of NGC~1637, we derive Hubble's constant for
   both the EPM and the promising SCM technique recently proposed by
   \cite{Hamuy02}: $H_0{\rm (EPM)\ } = 57 \pm 7 \pm 13\ \kmsmpc$ and $H_0{\rm
   (SCM)\ } = 59 \pm 3 \pm 11\ \kmsmpc$.  The uncertainty in both
   determinations of $H_0$ can be significantly reduced by obtaining
   additional calibrating galaxies.

5. We analyze the photometric behavior of SN~1999em from days $679\ {\rm to} \
   738$ after explosion.  After correcting for the effects of reddening and
   distance, SN~1999em and SN~1987A are found to have similar absolute $V$ and
   $I$ magnitudes at this phase.  From the comparison, we conclude that
   SN~1999em ejected about $0.09\ M_\odot$ of radioactive $^{56}{\rm Ni}$,
   which is slightly more than was derived for SN~1987A ($\sim 0.075\
   M_\odot$).  The $V$ and $I$ light curves of SN~1999em and SN~1987A are also
   quite similar, although SN~1999em may be fading somewhat more rapidly in the
   $I$ band.  Both objects decline at greater rates than the decay slope of
   $^{56}{\rm Co}$ predicts.  We postulate that increased transparency of the
   envelope to gamma rays, along with dust formation in the cooling atmosphere
   of the supernova, may explain this behavior.

\acknowledgments 

This research has made use of the NASA/IPAC Extragalactic Database (NED), which
is operated by the Jet Propulsion Laboratory, California Institute of
Technology, under contract with NASA.  We thank Dmitrij Nadyozhin, Tim Young,
Mario Hamuy, and Alex Filippenko for useful discussions, and an anonymous
referee for helpful suggestions that resulted in an improved manuscript.
Support for program \#G0-9155 was provided by NASA through a grant from the
Space Telescope Science Institute, which is operated by the Association of
Universities for Research in Astronomy, Inc., under NASA contract NAS 5-26555.

\clearpage


\clearpage
\begin{figure}
\plotone{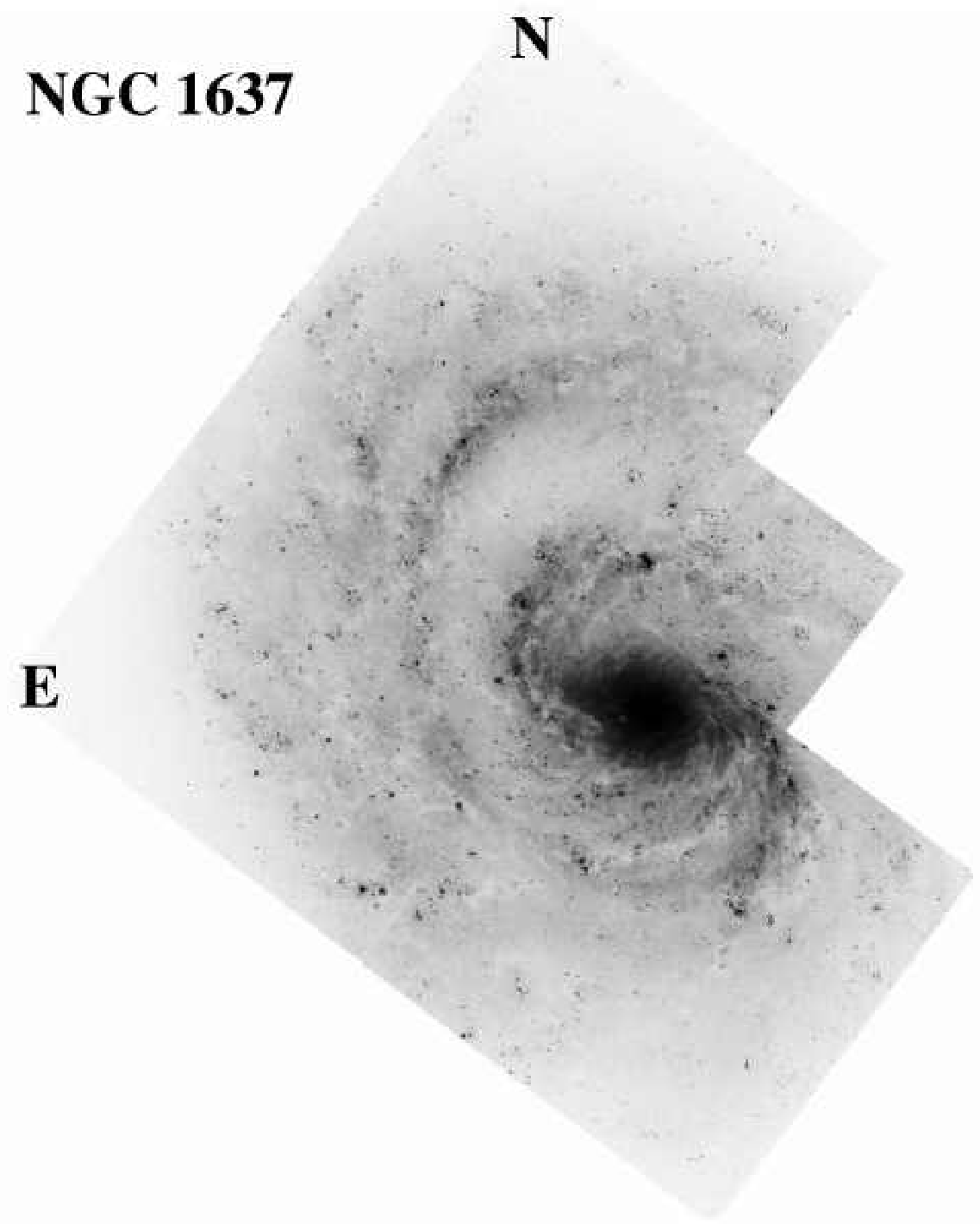}
\vskip -2.0in
\caption{Mosaic of the NGC~1637 field produced by co-adding all {\it HST} WFPC2
images obtained with the F555W filter ($V$-band).  The image has been rotated
so that North is up and East is to the left, as indicated.  The PC chip covers
the smallest field (chip 1).  Moving counterclockwise, the other three WF2
fields correspond to chips 2, 3, and 4.
\label{fig:1} }
\end{figure}

\begin{figure}
\plotone{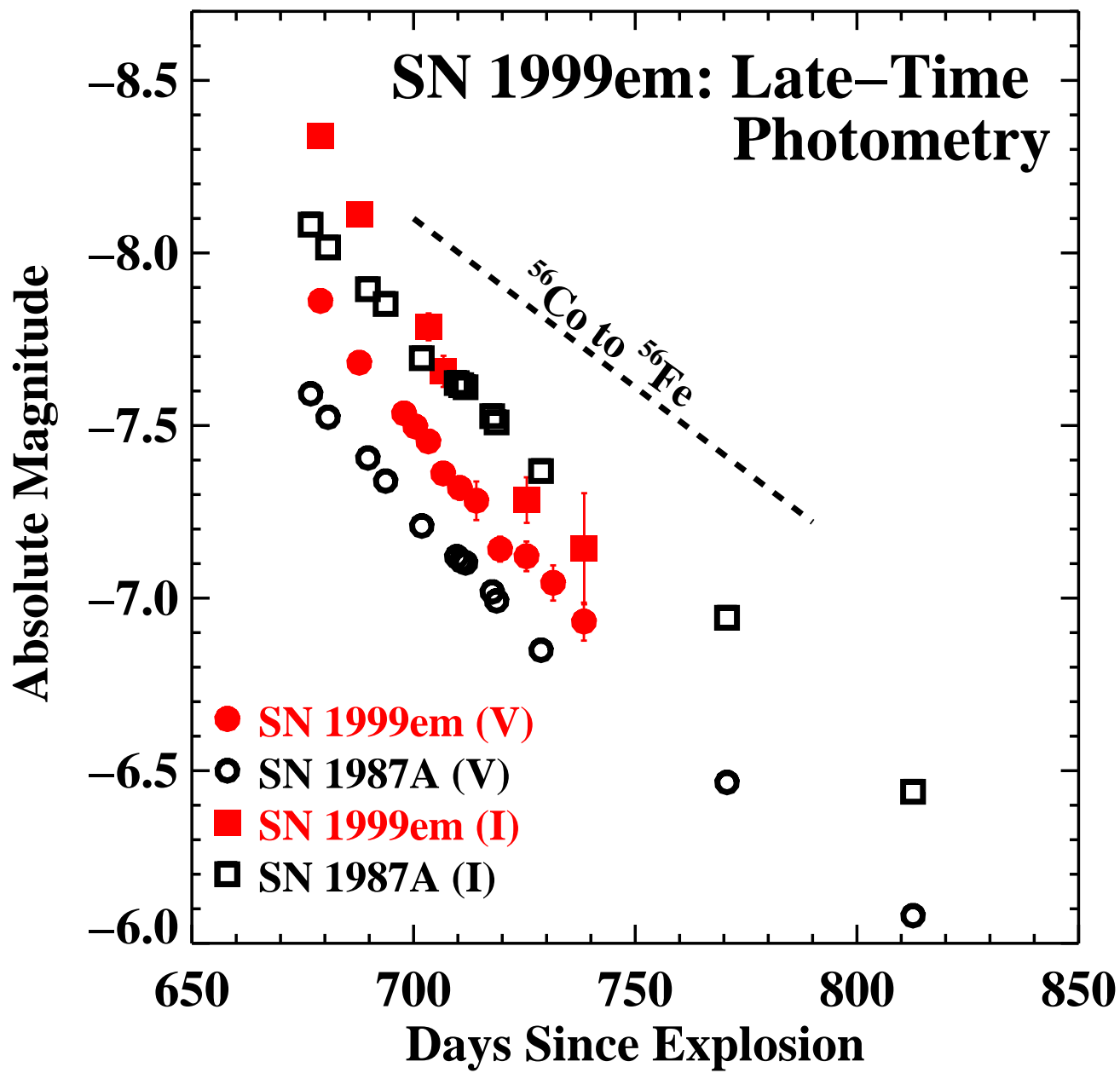}
\vskip 0.5in
\caption{Photometric behavior of SN~1999em ({\it closed, red symbols}) compared
with SN~1987A ({\it open, black symbols}) at $\sim 2$ years after explosion.
The fact that the optical and near-infrared brightnesses of both SNe decline
substantially faster than the rate predicted by the radioactive decay of $^{56}
{\rm Co}$ ({\it dashed line}) is attributed to increased envelope transparency
to gamma rays and to dust formation in the cooling SN atmospheres; see
text for details.
\label{fig:plot13} }
\end{figure}
\clearpage

\begin{figure}
\vskip -0.8in
\hskip -0.7in
\scalebox{1.2}{
\plotone{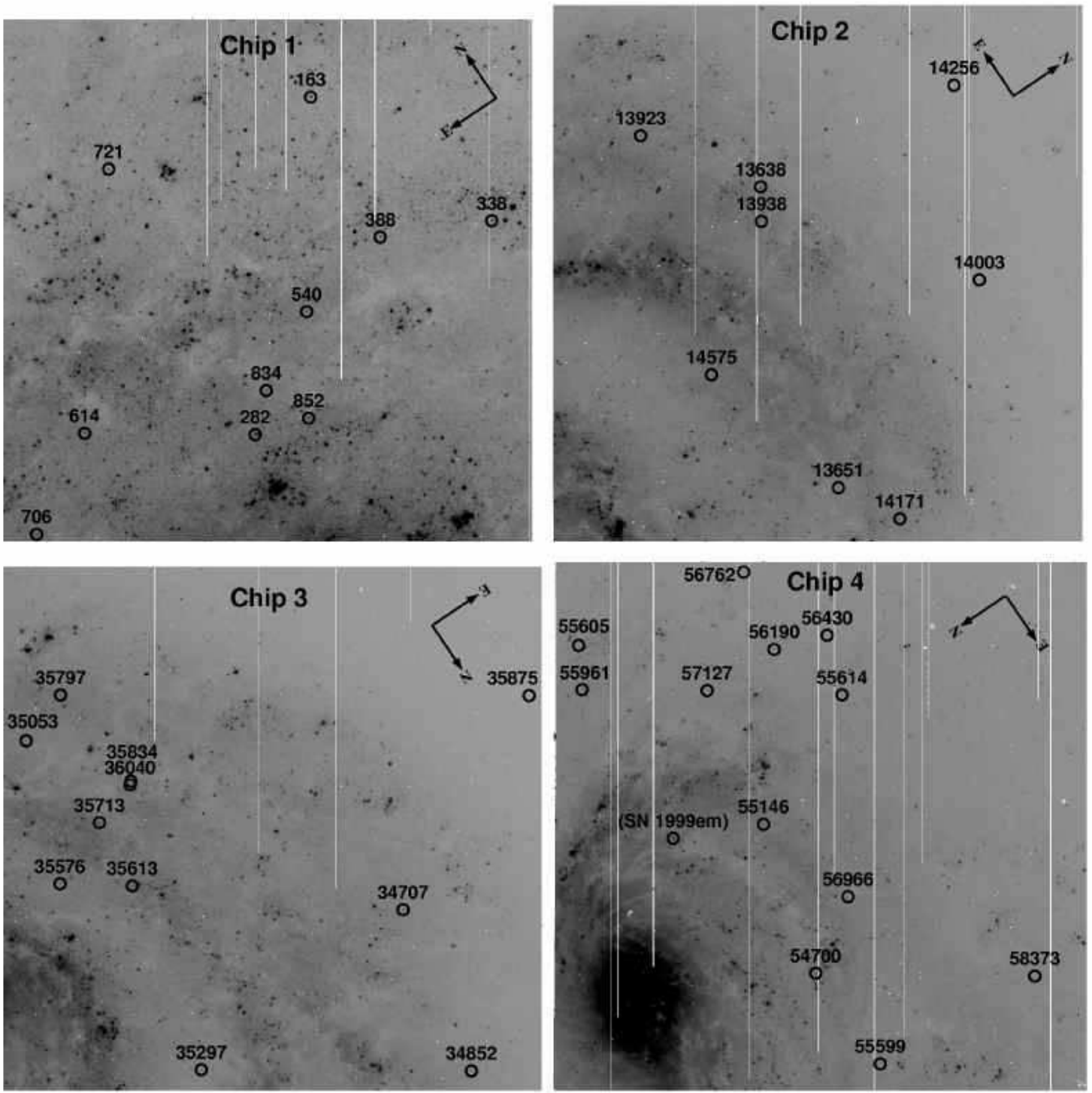}
}
\vskip -1.5in
\caption{Finding charts for the NGC~1637 Cepheids on the four WFPC2 chips, from
the combined $V$-band image.  The Cepheids are circled and labeled with their
identification number from Table~\ref{tab:tab6}.  Masked pixels are white.  The
field of view is $35\arcsec \times 35\arcsec$ for chip 1 (Planetary Camera of
the WFPC2), and $1\farcm3 \times 1\farcm3$ chips 2, 3, and 4 (Wide Field
Camera).  SN~1999em is also labeled on chip 4 ({\it hstphot} object number
54475, located at pixel position X=213.1, Y=407.8).
\label{fig:2} }
\end{figure}
\clearpage

\begin{figure}
\plotone{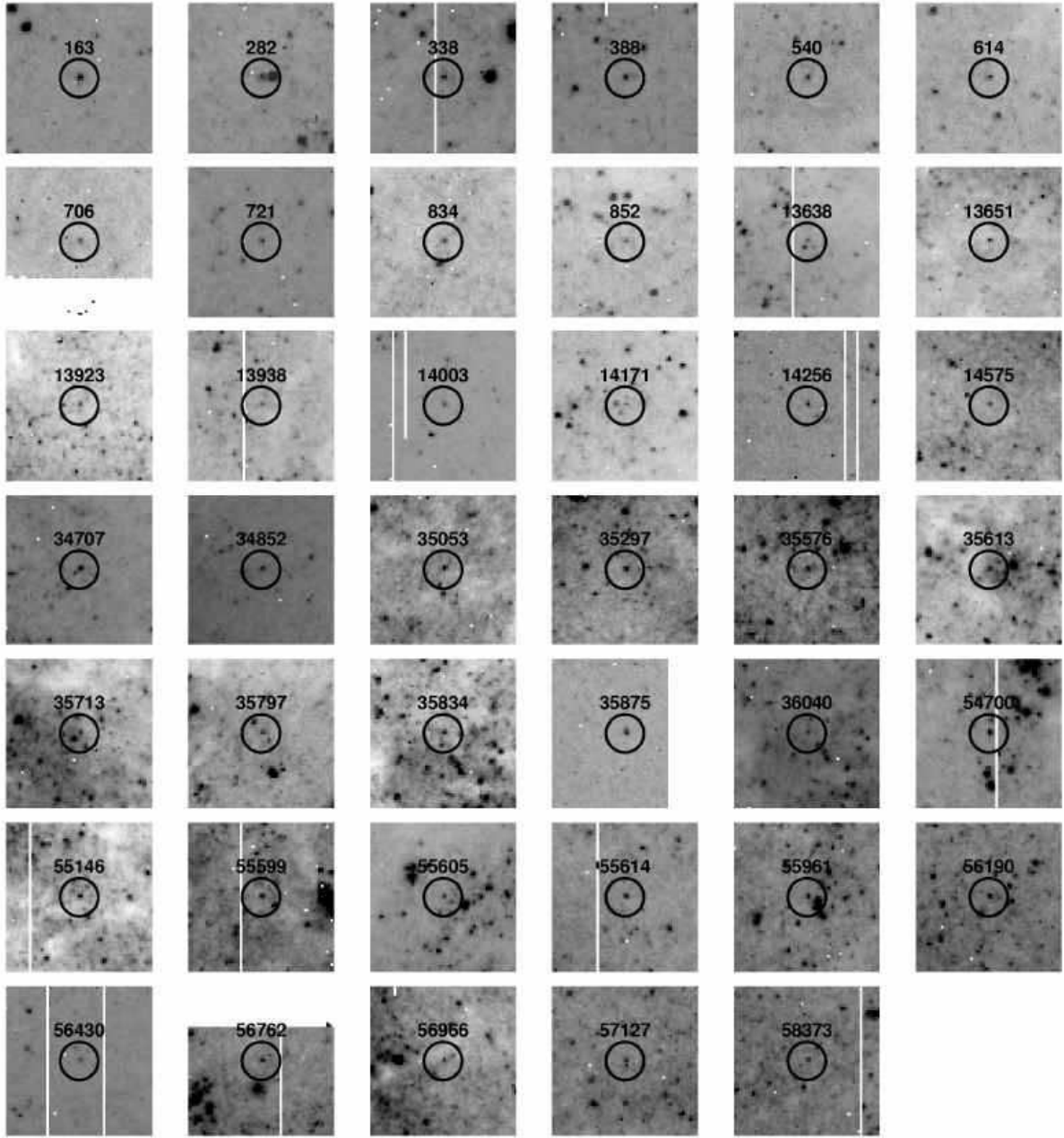}
\vskip -1.0in
\caption{Close-up finder charts for the NGC~1637 Cepheids, with the contrast
individually scaled to clearly show the candidate.  Each panel is a $60 \times
60$ pixel field of view centered on the variable.
\label{fig:3} }
\end{figure}
\clearpage

\begin{figure}
\scalebox{0.9}{
}
\caption{Light curves of candidate Cepheids folded at the most likely period.
The relative likelihoods (\S~\ref{sec:ioccv}) of the best-fitting template
light curves for periods between 10 and 100 days are also shown.
\label{fig:4} }
\end{figure}
\clearpage









\begin{figure}
\scalebox{0.85}{
\plotone{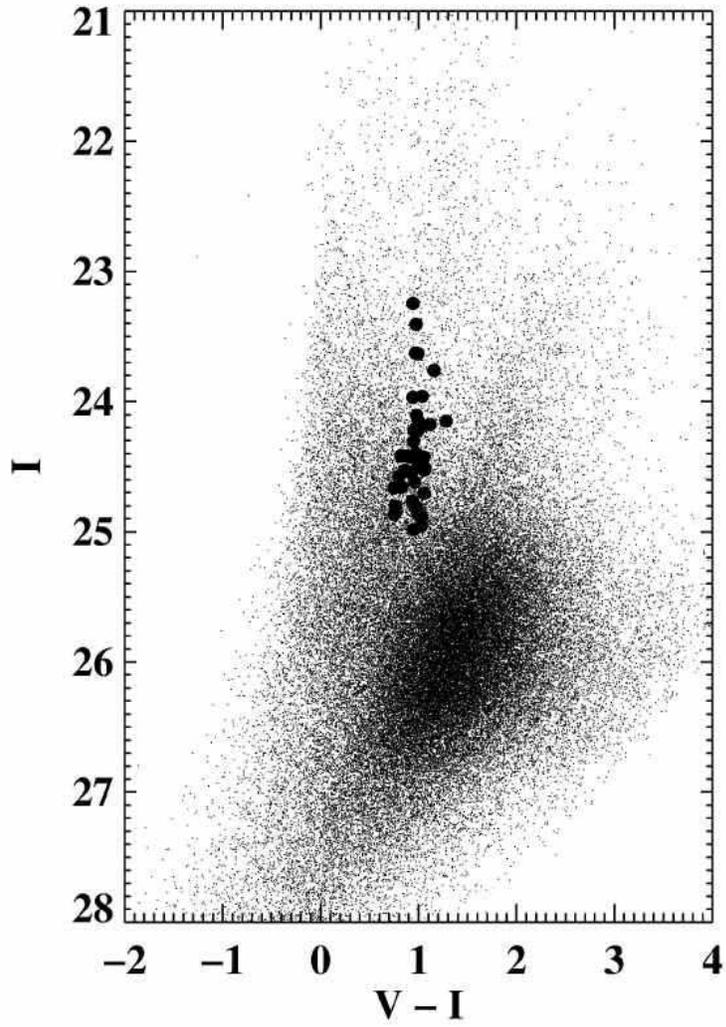}
}
\caption{Color-magnitude diagram of all of the stars in the NGC~1637 field,
with Cepheids indicated by {\it large closed circles}.  
\label{fig:6} }
\end{figure}

\begin{figure}
\vskip -1.5in
\plotone{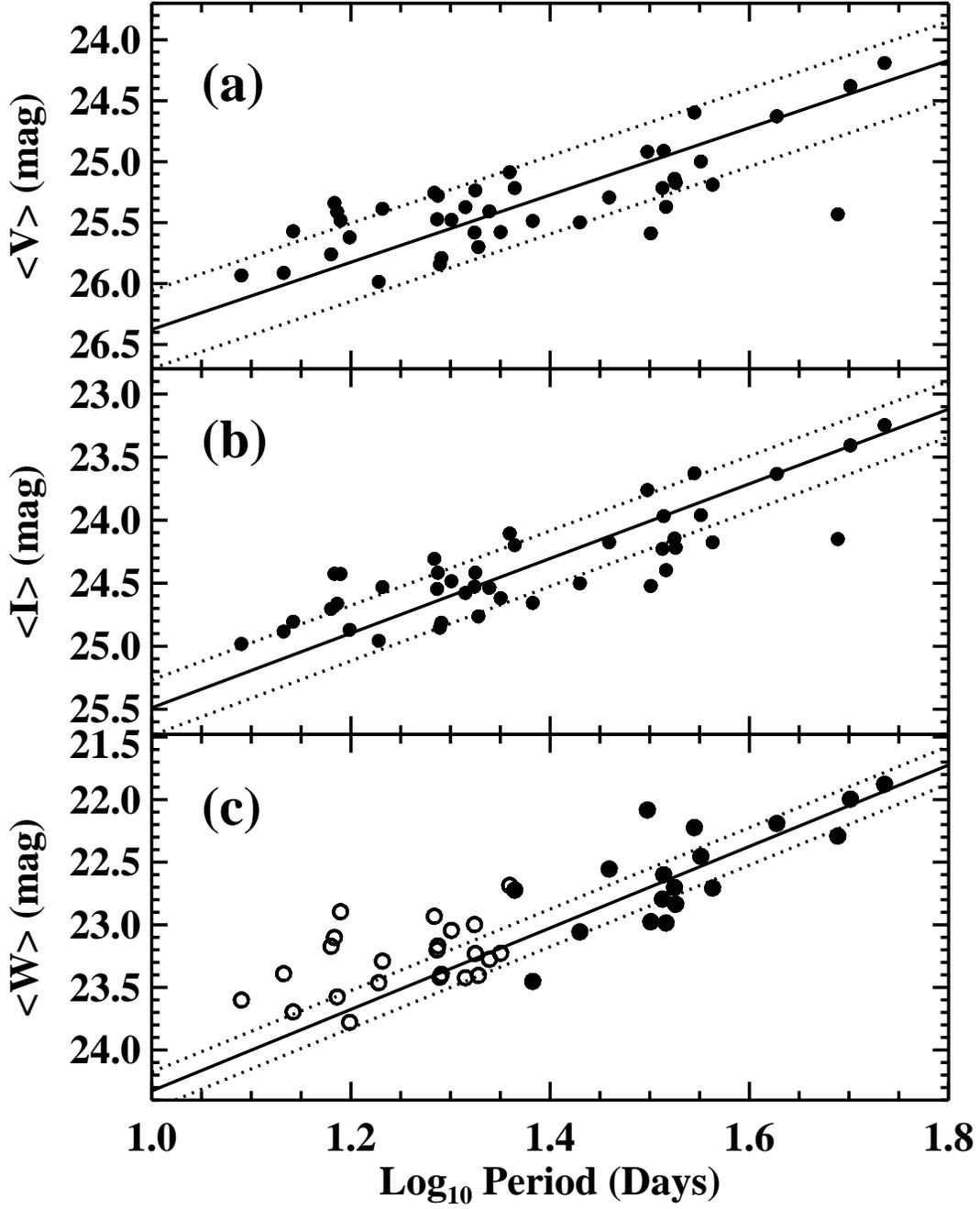}
\vskip -0.3in
\caption{({\it a}) Apparent $V$-band PL relation for the 41 Cepheids listed in
Table~\ref{tab:tab8}. The solid line shows a least-squares fit to all data
points, with the slope fixed to be that of the LMC PL relation adopted by the
KP (eq.~[\ref{eqn:twentyone}]) .  Dotted lines represent the expected intrinsic
$2\ \sigma$ scatter around the best fitting PL relation derived from the OGLE
LMC Cepheids. ({\it b}) As in ({\it a}), except for the $I$-band.  ({\it c}) As
in ({\it a}), except for the ``reddening-free'' $W$-band, with Cepheids
eliminated by the lower period cut described in the text indicated by {\it open
circles}.  The solid line shows a least-squares fit to the data points
represented by the {\it closed circles}, with the slope fixed to be that of the
LMC PL relation adopted by the KP (eq.~[\ref{eqn:twentytwo}]).
\label{fig:7} }
\end{figure}

\clearpage
\begin{figure}
\vskip -1.0in
\plotone{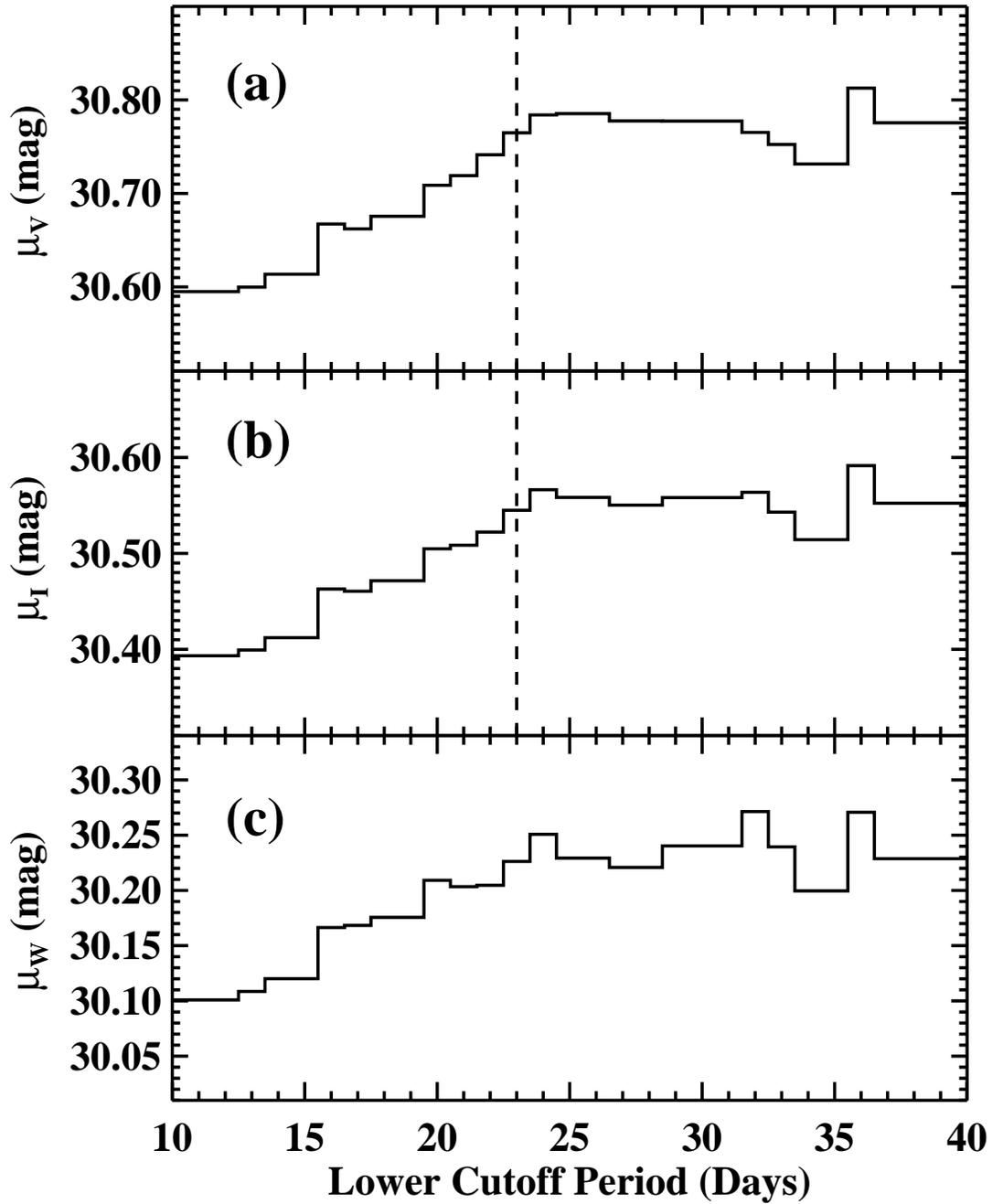}
\caption{Distance modulus as a function of lower cutoff period for ({\it
a}) $V$-band, ({\it b}) $I$-band, and ({\it c}) $W$-band.  The {\it vertical
dashed lines} in ({\it a}) and ({\it b}) indicate the point at which the
apparent distance moduli stabilize, thereby defining the lower period cutoff
threshold of $23.0$ days.  See text for details.
\label{fig:8} }
\end{figure}

\clearpage
\begin{figure}
\scalebox{0.9}{
\plotone{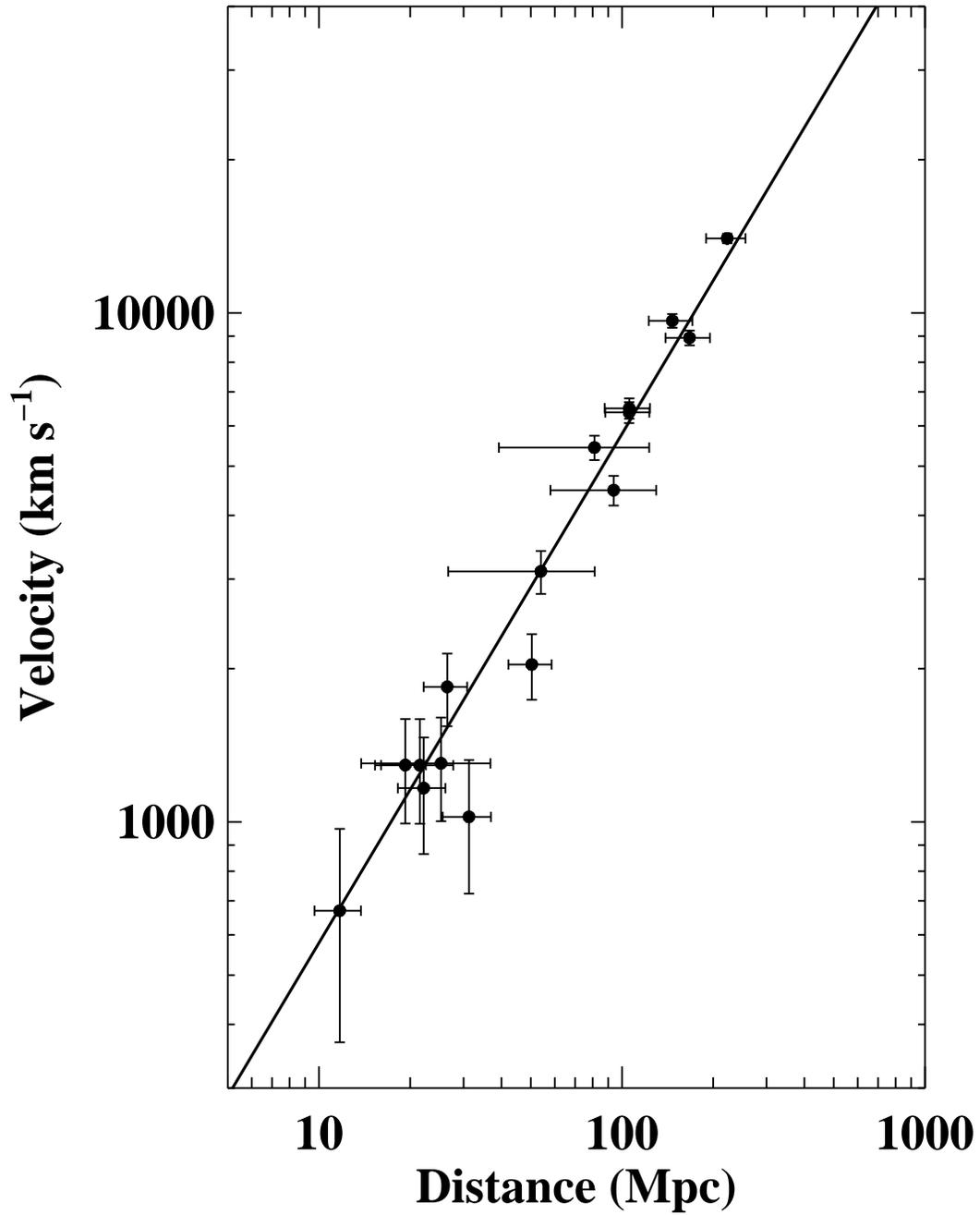}
}
\caption{$V$-band Hubble diagram for the standard-candle method of
\citet{Hamuy02}, as calibrated by the Cepheid distance to NGC~1637 derived in
this paper.
\label{fig:9} }
\end{figure}

\input{tab1.tex}

\input{tab2.tex}
\input{tab3.tex}
\input{tab4.tex}
\input{tab5.tex}
\input{tab6.tex}
\input{tab7.tex}
\input{tab8.tex}
\input{tab9.tex}

\end{document}

%% file: tab1.tex

\clearpage
\begin{deluxetable}{lrrrrrrl}
\tabletypesize{\small}
\tablenum{1}
\tablewidth{0pt}
\tablecaption{Slopes and Zero-Points for Cepheid PL Relations}
\tablehead{\colhead{Set} &
\colhead{$a_V$} &
\colhead{$a_I$} &
\colhead{$a_W$} &
\colhead{$b_V$}  &
\colhead{$b_I$} &
\colhead{$b_W$} &
\colhead{Reference}}

\startdata
A & $-2.760$ & $-2.962$ & $-3.255$ & $-1.458$ & $-1.942$ & $ -2.644$ & 1\\
B & $-2.760$ & $-3.060$ & $-3.495$ & $-1.400$ & $-1.810$ & $ -2.405$ & 2\\
C & $-2.480$ & $-2.820$ & $-3.313$ & $-1.750$ & $-2.090$ & $ -2.583$ & 3\\
D & $-3.140$ & $-3.410$ & $-3.802$ & $-0.830$ & $-1.330$ & $ -2.055$ & 3\\
\enddata

\tablecomments{Slopes and zero points for the PL relations resulting from the
four samples of Cepheids described in \S~\ref{sec:cvatplr}.  The relations are 
defined by $M_C = a_C \log (P) + b_C$, with $M_C$
the absolute magnitude of a Cepheid in photometric band $C$, $P$ its period
(usually measured in days), and $a_C$ and $b_C$ the slope and zero-point,
respectively, of the relation. $W$ represents the Weisenheit reddening-free
index described in the text.}   

\tablerefs{ (1) F01; (2) \citealt{Madore91}; (3) \citealt{Thim03}.}

\label{tab:tab1}

\end{deluxetable}
\clearpage 

%% file: tab2.tex
\def\ssp{\def\baselinestretch{1.0}\large\normalsize}
\begin{deluxetable}{lr}
\tabletypesize{\small}
\tablenum{2}
\tablewidth{0pt}
\tablecaption{Sources of Systematic Uncertainty in the Cepheid Distance to NGC 1637}
\tablehead{\colhead{} &
\colhead{Error} \\
\colhead{Source of Uncertainty} &
\colhead{(mag)} }

\startdata
Zero-point of PL relations & $0.10$ \\
Metallicity                & $0.12$ \\
Slope of PL relations      & $0.06$ \\
Aperture corrections       & $0.02$ \\
WFPC2 zero-point           & $0.04$ \\
\cline{2-2} 
Total systematic uncertainty & $0.17$ \\

\enddata

\label{tab:tab2}
\end{deluxetable}

\clearpage


%% file: tab3.tex

\clearpage
\begin{deluxetable}{llccrc}
\small
\tablenum{3}
\tablecaption{{\it HST} Observations of NGC 1637}
\tablehead{\colhead{} &
\colhead{Data Archive} &
\colhead{} &
\colhead{HJD at}  &
\colhead{Relative } &
\colhead{Exposure Time}\\
\colhead{Epoch} &
\colhead{Designation} &
\colhead{UT Date} &
\colhead{Midexposure}  &
\colhead{Day\tablenotemark{a}} &
\colhead{(s)} \\
\hline\\
\multicolumn{6}{c}{F555W} }

\startdata

1  & u6fv0101/2m & 2001 Sep 02 & 2452154.637 & 0.000  & 1100 + 1100 \\
2  & u6fv0201/2r & 2001 Sep 10 & 2452163.395 & 8.759  & 1100 + 1100 \\
3  & u6fv0301/2m & 2001 Sep 20 & 2452173.494 & 18.857 & 1100 + 1100 \\
4  & u6fv0401/2m & 2001 Sep 23 & 2452176.029 & 21.392 & 1100 + 1100 \\
5  & u6fv0501/2m & 2001 Sep 26 & 2452178.975 & 24.338 & 1100 + 1100 \\
6  & u6fv0601/2m & 2001 Sep 29 & 2452182.319 & 27.682 & 1100 + 1100 \\
7  & u6fv0701/2m & 2001 Oct 03 & 2452186.061 & 31.425 & 1100 + 1100 \\
8  & u6fv0801/2m & 2001 Oct 07 & 2452189.804 & 35.167 & 1100 + 1100 \\
9  & u6fv0901/2m & 2001 Oct 12 & 2452195.152 & 40.515 & 1100 + 1100 \\
10 & u6fv1001/2m & 2001 Oct 18 & 2452201.105 & 46.469 & 1100 + 1100 \\
11 & u6fv1101/2m & 2001 Oct 24 & 2452207.117 & 52.481 & 1100 + 1100 \\
12 & u6fv1201/2m & 2001 Oct 31 & 2452214.071 & 59.434 & 1100 + 1100 \\

\hline\\
\multicolumn{6}{c}{F814W} \\
\hline\\

1  & u6fv0103/4m & 2001 Sep 02 & 2452154.699 & 0.000  & 1100 + 1100 \\
2  & u6fv0203/4r & 2001 Sep 10 & 2452163.458 & 8.759  & 1100 + 1100 \\
5  & u6fv0503/4m & 2001 Sep 26 & 2452179.041 & 24.342 & 1100 + 1100 \\
6  & u6fv0603/4m & 2001 Sep 29 & 2452182.381 & 27.682 & 1100 + 1100 \\
10 & u6fv1003/4m & 2001 Oct 18 & 2452201.168 & 46.469 & 1100 + 1100 \\
12 & u6fv1203/4m & 2001 Oct 31 & 2452214.137 & 59.438 & 1100 + 1100 \\

\enddata

\tablenotetext{a}{Day since first epoch in each filter.}

\label{tab:tab3}
\end{deluxetable}

\clearpage


%% file: tab4.tex

\ssp
\clearpage
\begin{deluxetable}{rcrrcc}
\tabletypesize{\footnotesize}
\tablenum{4}
\tablewidth{0pt}
\tablecaption{Reference Star Photometry}
\tablehead{\colhead{ID} &
\colhead{Chip} &
\colhead{$X$} &
\colhead{$Y$} &
\colhead{$V$}  &
\colhead{$I$}}

\startdata
$   16$ & $1$ & $507.7$ & $195.9$ & $22.87\pm0.00$ & $21.95\pm0.01$ \\
$   49$ & $1$ & $170.6$ & $700.8$ & $23.50\pm0.01$ & $23.07\pm0.01$ \\
$   52$ & $1$ & $599.6$ & $120.7$ & $23.50\pm0.01$ & $23.22\pm0.02$ \\
$   94$ & $1$ & $258.4$ & $543.1$ & $23.91\pm0.01$ & $23.78\pm0.02$ \\
$12912$ & $2$ & $117.1$ & $ 64.4$ & $22.72\pm0.00$ & $22.44\pm0.01$ \\
$12929$ & $2$ & $196.8$ & $717.9$ & $22.82\pm0.00$ & $22.79\pm0.01$ \\
$12932$ & $2$ & $331.0$ & $583.5$ & $23.09\pm0.00$ & $22.55\pm0.01$ \\
$12935$ & $2$ & $226.1$ & $384.5$ & $23.00\pm0.00$ & $22.97\pm0.01$ \\
$12955$ & $2$ & $514.1$ & $196.7$ & $23.39\pm0.01$ & $22.78\pm0.01$ \\
$12967$ & $2$ & $270.9$ & $ 69.3$ & $23.53\pm0.01$ & $22.83\pm0.01$ \\
$13022$ & $2$ & $352.9$ & $149.2$ & $23.88\pm0.01$ & $23.38\pm0.02$ \\
$13035$ & $2$ & $373.8$ & $291.3$ & $23.95\pm0.01$ & $23.48\pm0.02$ \\
$34397$ & $3$ & $257.5$ & $154.6$ & $22.85\pm0.00$ & $22.55\pm0.01$ \\
$34410$ & $3$ & $241.7$ & $607.9$ & $23.08\pm0.00$ & $22.38\pm0.01$ \\
$34420$ & $3$ & $210.6$ & $310.5$ & $23.37\pm0.01$ & $21.96\pm0.01$ \\
$34455$ & $3$ & $397.0$ & $258.3$ & $23.23\pm0.01$ & $23.27\pm0.02$ \\
$34493$ & $3$ & $231.3$ & $568.2$ & $23.59\pm0.01$ & $23.47\pm0.02$ \\
$34510$ & $3$ & $665.5$ & $332.5$ & $23.63\pm0.01$ & $23.78\pm0.03$ \\
$34547$ & $3$ & $359.8$ & $569.8$ & $23.69\pm0.01$ & $23.74\pm0.02$ \\
$34586$ & $3$ & $278.7$ & $339.1$ & $23.97\pm0.01$ & $23.70\pm0.03$ \\
$34598$ & $3$ & $145.4$ & $244.5$ & $23.93\pm0.01$ & $23.84\pm0.03$ \\
$54399$ & $4$ & $545.7$ & $680.5$ & $22.24\pm0.00$ & $21.63\pm0.01$ \\
$54445$ & $4$ & $121.4$ & $526.4$ & $23.25\pm0.00$ & $21.68\pm0.00$ \\
$54478$ & $4$ & $330.6$ & $227.4$ & $23.13\pm0.00$ & $23.02\pm0.01$ \\
$54506$ & $4$ & $166.6$ & $681.4$ & $23.41\pm0.01$ & $22.88\pm0.01$ \\
$54516$ & $4$ & $290.1$ & $ 87.2$ & $23.67\pm0.01$ & $22.34\pm0.01$ \\
$54560$ & $4$ & $208.2$ & $313.0$ & $23.65\pm0.01$ & $23.25\pm0.02$ \\
$54567$ & $4$ & $671.8$ & $284.6$ & $23.61\pm0.01$ & $23.40\pm0.02$ \\
$54587$ & $4$ & $127.3$ & $471.3$ & $23.75\pm0.01$ & $23.40\pm0.03$ \\
$54594$ & $4$ & $247.5$ & $488.5$ & $23.76\pm0.01$ & $23.22\pm0.02$ \\
$54598$ & $4$ & $374.3$ & $184.1$ & $23.77\pm0.01$ & $23.28\pm0.02$ \\
$54626$ & $4$ & $350.5$ & $555.4$ & $23.81\pm0.01$ & $23.55\pm0.02$ \\
$54641$ & $4$ & $ 73.5$ & $419.8$ & $23.94\pm0.01$ & $23.69\pm0.03$ \\
$54684$ & $4$ & $766.5$ & $379.5$ & $23.97\pm0.01$ & $23.70\pm0.02$ \\
$54692$ & $4$ & $ 66.6$ & $692.6$ & $23.94\pm0.01$ & $24.01\pm0.03$ \\
$54709$ & $4$ & $396.2$ & $416.7$ & $23.96\pm0.01$ & $23.89\pm0.03$ \\
$54725$ & $4$ & $507.7$ & $268.3$ & $23.96\pm0.01$ & $24.12\pm0.03$ \\

\enddata 

\tablecomments{Photometry of bright, non-variable stars in NGC~1637.
Uncertainties are the formal photometric errors reported by {\it hstphot} for
measurements made on the combined $V$ (12 epochs) and $I$ (6 epochs) frames.
Note that formal uncertainties that are less than $0.005$ mag are reported here
as $0.00$ mag.}

\label{tab:tab4}

\end{deluxetable}

\clearpage


%% file: tab5.tex

\clearpage          
\begin{deluxetable}{lccc}
\tablenum{5}
\tablewidth{0pt}
\tablecaption{Photometry of SN 1999em}
\tablehead{\colhead{Epoch\tablenotemark{a}}  &
\colhead{Day\tablenotemark{b}} &
\colhead{$V$ ($\sigma_V$)} &
\colhead{$I$ ($\sigma_I$)} }

\startdata

1   &  679 & 22.791    (0.016) &        22.187    (0.021) \nl
2   &  688 & 22.970    (0.017) &        22.415    (0.023)\nl
3   &  698 & 23.117    (0.022) & \nodata\nl
4   &  700 & 23.155    (0.027) & \nodata\nl
5   &  703 & 23.198    (0.030) &        22.740    (0.039)\nl
6   &  707 & 23.291    (0.027) &        22.869    (0.045)\nl
7   &  710 & 23.333    (0.034) & \nodata\nl
8   &  714 & 23.370    (0.056) & \nodata\nl
9   &  720 & 23.510    (0.036) & \nodata\nl
10  &  725 & 23.531    (0.043) &        23.242    (0.066)\nl
11  &  731 & 23.608    (0.051) & \nodata\nl
12  &  738 & 23.720    (0.055) &        23.383     (0.161)\nl

\enddata

\tablenotetext{a}{As defined by Table~\ref{tab:tab3}.}
\tablenotetext{b}{Day since estimated explosion date of HJD 2,451,475.64
(Leonard et al. 2002a).}

\label{tab:tab5}

\end{deluxetable}

\clearpage


%% file: tab6.tex

\clearpage
\begin{deluxetable}{lcrr}
\tabletypesize{\scriptsize}
\tablenum{6}
\tablewidth{0pt}
\tablecaption{Coordinates of Cepheid Variables in NGC 1637}
\tablehead{\colhead{ID} &
\colhead{Chip} &
\colhead{$X$} &
\colhead{$Y$} }

\startdata
$  163$ & $1$ & $485.9$ & $688.5$ \\
$  282$ & $1$ & $407.6$ & $210.8$ \\
$  338$ & $1$ & $742.1$ & $513.5$ \\
$  388$ & $1$ & $584.5$ & $490.2$ \\
$  540$ & $1$ & $480.4$ & $384.6$ \\
$  614$ & $1$ & $166.2$ & $212.5$ \\
$  706$ & $1$ & $ 98.2$ & $ 69.3$ \\
$  721$ & $1$ & $200.2$ & $586.6$ \\
$  834$ & $1$ & $423.3$ & $273.0$ \\
$  852$ & $1$ & $483.2$ & $233.8$ \\
$13638$ & $2$ & $342.8$ & $533.3$ \\
$13651$ & $2$ & $452.9$ & $107.3$ \\
$13923$ & $2$ & $173.2$ & $605.8$ \\
$13938$ & $2$ & $344.0$ & $484.7$ \\
$14003$ & $2$ & $652.3$ & $401.9$ \\
$14171$ & $2$ & $540.0$ & $ 63.3$ \\
$14256$ & $2$ & $616.5$ & $677.3$ \\
$14575$ & $2$ & $273.3$ & $267.3$ \\
$34707$ & $3$ & $602.7$ & $312.7$ \\
$34852$ & $3$ & $699.3$ & $ 84.4$ \\
$35053$ & $3$ & $ 68.4$ & $550.8$ \\
$35297$ & $3$ & $317.2$ & $ 86.0$ \\
$35576$ & $3$ & $117.2$ & $349.2$ \\
$35613$ & $3$ & $219.7$ & $346.3$ \\
$35713$ & $3$ & $173.1$ & $435.9$ \\
$35797$ & $3$ & $117.9$ & $616.8$ \\
$35834$ & $3$ & $217.7$ & $494.6$ \\
$35875$ & $3$ & $781.0$ & $615.9$ \\
$36040$ & $3$ & $216.5$ & $489.3$ \\
$54700$ & $4$ & $414.8$ & $216.8$ \\
$55146$ & $4$ & $341.1$ & $427.5$ \\
$55599$ & $4$ & $506.2$ & $ 88.4$ \\
$55605$ & $4$ & $ 79.3$ & $680.9$ \\
$55614$ & $4$ & $452.2$ & $610.9$ \\
$55961$ & $4$ & $ 84.4$ & $618.1$ \\
$56190$ & $4$ & $356.1$ & $675.1$ \\
$56430$ & $4$ & $430.9$ & $695.4$ \\
$56762$ & $4$ & $313.0$ & $784.7$ \\
$56966$ & $4$ & $460.6$ & $325.1$ \\
$57127$ & $4$ & $261.3$ & $616.8$ \\
$58373$ & $4$ & $724.9$ & $212.8$ \\

\enddata
\label{tab:tab6}

\end{deluxetable}

\clearpage


%% file: tab7.tex

\clearpage
\begin{deluxetable}{lrrrrrrrrrrr}
\tabletypesize{\tiny}
\tablenum{7}
\tablewidth{0pt}
\tablecaption{Photometry of Cepheid Variables in NGC 1637}
\tablehead{
\colhead{Epoch} &
\colhead{$V$}  &
\colhead{$I$} &
\colhead{} &
\colhead{$V$} &
\colhead{$I$} &
\colhead{} &
\colhead{$V$} &
\colhead{$I$} &
\colhead{} &
\colhead{$V$}  &
\colhead{$I$}}

\startdata

\hline\\
\colhead{} &
\multicolumn{2}{c}{  163}  &
\colhead{} &
\multicolumn{2}{c}{  282}  &
\colhead{} &
\multicolumn{2}{c}{  338}  &
\colhead{} &
\multicolumn{2}{c}{  388} \\
\cline{2-3} \cline{5-6} \cline{8-9} \cline{11-12}
$ 1$ & $24.83\pm0.08$ & $23.82\pm0.08$ & & $24.68\pm0.07$ & $23.77\pm0.08$ & & $24.86\pm0.06$ & $23.92\pm0.06$ & & $24.82\pm0.09$ & $23.93\pm0.07$    \\
$ 2$ & $25.13\pm0.08$ & $24.00\pm0.06$ & & $25.01\pm0.08$ & $23.90\pm0.07$ & & $24.93\pm0.07$ & $23.97\pm0.06$ & & $25.33\pm0.08$ & $24.22\pm0.08$    \\
$ 3$ & $24.10\pm0.03$ & \nodata & & $25.58\pm0.10$ & \nodata & & $25.42\pm0.10$ & \nodata & & $25.72\pm0.12$ & \nodata    \\
$ 4$ & $24.28\pm0.05$ & \nodata & & $25.55\pm0.11$ & \nodata & & $25.64\pm0.12$ & \nodata & & $25.63\pm0.10$ & \nodata    \\
$ 5$ & $24.33\pm0.05$ & $23.43\pm0.08$ & & $25.31\pm0.08$ & $24.34\pm0.09$ & & $25.53\pm0.10$ & $24.47\pm0.10$ & & $25.75\pm0.10$ & $24.63\pm0.10$    \\
$ 6$ & $24.55\pm0.05$ & $23.50\pm0.04$ & & $24.91\pm0.06$ & $23.90\pm0.08$ & & $25.49\pm0.11$ & $24.54\pm0.10$ & & $25.04\pm0.06$ & $24.18\pm0.08$    \\
$ 7$ & $24.74\pm0.05$ & \nodata & & $24.49\pm0.05$ & \nodata & & $25.39\pm0.09$ & \nodata & & $24.61\pm0.06$ & \nodata    \\
$ 8$ & $24.78\pm0.09$ & \nodata & & $24.65\pm0.06$ & \nodata & & $24.59\pm0.04$ & \nodata & & $24.99\pm0.06$ & \nodata    \\
$ 9$ & $25.03\pm0.06$ & \nodata & & $24.93\pm0.06$ & \nodata & & $24.81\pm0.05$ & \nodata & & $25.05\pm0.13$ & \nodata    \\
$10$ & $25.14\pm0.07$ & $23.94\pm0.10$ & & $25.15\pm0.10$ & $24.01\pm0.09$ & & $25.18\pm0.07$ & $24.02\pm0.10$ & & $25.22\pm0.12$ & $24.16\pm0.07$    \\
$11$ & $24.02\pm0.04$ & \nodata & & $25.32\pm0.14$ & \nodata & & $25.36\pm0.08$ & \nodata & & $25.58\pm0.10$ & \nodata    \\
$12$ & $24.35\pm0.04$ & $23.45\pm0.06$ & & $25.47\pm0.11$ & $24.33\pm0.11$ & & $25.68\pm0.13$ & $24.63\pm0.11$ & & $25.62\pm0.10$ & $24.41\pm0.11$    \\
\hline\\
\colhead{} &
\multicolumn{2}{c}{  540}  &
\colhead{} &
\multicolumn{2}{c}{  614}  &
\colhead{} &
\multicolumn{2}{c}{  706}  &
\colhead{} &
\multicolumn{2}{c}{  721} \\
\cline{2-3} \cline{5-6} \cline{8-9} \cline{11-12}
$ 1$ & $26.11\pm0.16$ & $25.19\pm0.19$ & & $25.65\pm0.10$ & $24.74\pm0.12$ & & $25.15\pm0.09$ & $24.68\pm0.16$ & & $26.10\pm0.15$ & $24.99\pm0.15$    \\
$ 2$ & $25.37\pm0.08$ & $24.27\pm0.08$ & & $25.87\pm0.11$ & $24.83\pm0.12$ & & $25.91\pm0.13$ & $24.84\pm0.15$ & & $25.31\pm0.07$ & $24.65\pm0.11$    \\
$ 3$ & $26.05\pm0.12$ & \nodata & & $25.42\pm0.09$ & \nodata & & $25.30\pm0.08$ & \nodata & & $25.98\pm0.12$ & \nodata    \\
$ 4$ & $26.00\pm0.16$ & \nodata & & $25.70\pm0.12$ & \nodata & & $25.45\pm0.10$ & \nodata & & $26.19\pm0.14$ & \nodata    \\
$ 5$ & $25.84\pm0.11$ & $25.05\pm0.17$ & & $25.77\pm0.15$ & $25.09\pm0.16$ & & $25.83\pm0.15$ & $24.93\pm0.13$ & & $26.10\pm0.15$ & $24.94\pm0.14$    \\
$ 6$ & $24.90\pm0.05$ & $24.35\pm0.09$ & & $25.70\pm0.11$ & $24.80\pm0.11$ & & $26.04\pm0.13$ & $24.66\pm0.13$ & & $25.62\pm0.11$ & $24.66\pm0.10$    \\
$ 7$ & $25.17\pm0.08$ & \nodata & & $25.35\pm0.09$ & \nodata & & $26.48\pm0.30$ & \nodata & & $25.29\pm0.07$ & \nodata    \\
$ 8$ & $25.40\pm0.09$ & \nodata & & $25.66\pm0.10$ & \nodata & & $26.11\pm0.14$ & \nodata & & $25.55\pm0.09$ & \nodata    \\
$ 9$ & $25.89\pm0.14$ & \nodata & & $25.80\pm0.10$ & \nodata & & $25.36\pm0.11$ & \nodata & & $26.14\pm0.13$ & \nodata    \\
$10$ & $25.99\pm0.13$ & $25.05\pm0.15$ & & $25.33\pm0.07$ & $24.74\pm0.11$ & & $25.88\pm0.14$ & $24.68\pm0.14$ & & $25.92\pm0.12$ & $25.11\pm0.19$    \\
$11$ & $24.78\pm0.05$ & \nodata & & $25.91\pm0.13$ & \nodata & & $26.63\pm0.25$ & \nodata & & $25.33\pm0.08$ & \nodata    \\
$12$ & $25.50\pm0.08$ & $24.58\pm0.14$ & & $25.32\pm0.07$ & $24.67\pm0.11$ & & $25.35\pm0.09$ & $24.47\pm0.12$ & & $25.77\pm0.10$ & $24.76\pm0.13$    \\
\hline\\
\colhead{} &
\multicolumn{2}{c}{  834}  &
\colhead{} &
\multicolumn{2}{c}{  852}  &
\colhead{} &
\multicolumn{2}{c}{13638}  &
\colhead{} &
\multicolumn{2}{c}{13651} \\
\cline{2-3} \cline{5-6} \cline{8-9} \cline{11-12}
$ 1$ & $26.72\pm0.26$ & $24.97\pm0.18$ & & $25.91\pm0.12$ & $24.79\pm0.15$ & & $25.00\pm0.09$ & $24.68\pm0.11$ & & $25.12\pm0.08$ & $23.98\pm0.06$    \\
$ 2$ & $25.80\pm0.15$ & $24.72\pm0.18$ & & $25.75\pm0.13$ & $24.94\pm0.18$ & & $25.41\pm0.13$ & $24.43\pm0.09$ & & $25.70\pm0.11$ & $24.49\pm0.12$    \\
$ 3$ & $25.34\pm0.07$ & \nodata & & $26.52\pm0.25$ & \nodata & & $25.46\pm0.09$ & \nodata & & $26.14\pm0.14$ & \nodata    \\
$ 4$ & $25.43\pm0.16$ & \nodata & & $25.44\pm0.08$ & \nodata & & $24.86\pm0.05$ & \nodata & & $24.75\pm0.06$ & \nodata    \\
$ 5$ & $25.99\pm0.12$ & $25.09\pm0.15$ & & $25.68\pm0.12$ & $24.66\pm0.14$ & & $25.02\pm0.07$ & $24.06\pm0.10$ & & $24.70\pm0.05$ & $23.84\pm0.06$    \\
$ 6$ & $26.88\pm0.26$ & $24.92\pm0.15$ & & $26.40\pm0.17$ & $25.14\pm0.19$ & & $25.27\pm0.08$ & $24.38\pm0.08$ & & $24.96\pm0.06$ & $23.86\pm0.16$    \\
$ 7$ & $26.04\pm0.12$ & \nodata & & $26.45\pm0.20$ & \nodata & & $25.66\pm0.12$ & \nodata & & $25.23\pm0.08$ & \nodata    \\
$ 8$ & $25.79\pm0.10$ & \nodata & & $25.45\pm0.09$ & \nodata & & $25.58\pm0.12$ & \nodata & & $25.57\pm0.11$ & \nodata    \\
$ 9$ & $26.23\pm0.14$ & \nodata & & $26.71\pm0.25$ & \nodata & & $24.67\pm0.05$ & \nodata & & $25.67\pm0.10$ & \nodata    \\
$10$ & $25.44\pm0.08$ & $24.66\pm0.11$ & & $25.34\pm0.08$ & $24.76\pm0.17$ & & $25.47\pm0.09$ & $24.39\pm0.12$ & & $25.93\pm0.19$ & $24.68\pm0.10$    \\
$11$ & $26.23\pm0.17$ & \nodata & & $26.47\pm0.21$ & \nodata & & $25.61\pm0.09$ & \nodata & & $24.75\pm0.05$ & \nodata    \\
$12$ & $25.44\pm0.08$ & $24.71\pm0.12$ & & $25.44\pm0.10$ & $24.83\pm0.22$ & & $24.80\pm0.05$ & $24.09\pm0.06$ & & $25.25\pm0.07$ & $23.98\pm0.06$    \\
\hline\\
\colhead{} &
\multicolumn{2}{c}{13923}  &
\colhead{} &
\multicolumn{2}{c}{13938}  &
\colhead{} &
\multicolumn{2}{c}{14003}  &
\colhead{} &
\multicolumn{2}{c}{14171} \\
\cline{2-3} \cline{5-6} \cline{8-9} \cline{11-12}
$ 1$ & $25.84\pm0.16$ & $24.73\pm0.32$ & & $25.43\pm0.09$ & $24.45\pm0.09$ & & $25.72\pm0.09$ & $24.64\pm0.11$ & & $25.61\pm0.18$ & $24.58\pm0.14$    \\
$ 2$ & $25.72\pm0.13$ & $24.53\pm0.10$ & & $25.75\pm0.17$ & $25.04\pm0.15$ & & $25.50\pm0.08$ & $24.49\pm0.09$ & & $26.56\pm0.39$ & $24.97\pm0.16$    \\
$ 3$ & $25.11\pm0.07$ & \nodata & & $24.75\pm0.09$ & \nodata & & $24.84\pm0.19$ & \nodata & & $25.16\pm0.09$ & \nodata    \\
$ 4$ & $25.37\pm0.09$ & \nodata & & $25.33\pm0.09$ & \nodata & & $25.22\pm0.06$ & \nodata & & $25.31\pm0.09$ & \nodata    \\
$ 5$ & $25.23\pm0.08$ & $24.36\pm0.09$ & & $25.39\pm0.29$ & $24.39\pm0.09$ & & $25.62\pm0.08$ & $24.38\pm0.07$ & & $25.62\pm0.11$ & $24.37\pm0.09$    \\
$ 6$ & $25.47\pm0.13$ & $24.38\pm0.09$ & & $26.11\pm0.37$ & $24.87\pm0.54$ & & $25.76\pm0.10$ & $24.90\pm0.11$ & & $25.86\pm0.13$ & $24.91\pm0.16$    \\
$ 7$ & $25.77\pm0.12$ & \nodata & & $25.74\pm0.12$ & \nodata & & $25.39\pm0.09$ & \nodata & & $25.64\pm0.16$ & \nodata    \\
$ 8$ & $26.31\pm0.20$ & \nodata & & $25.71\pm0.11$ & \nodata & & $24.99\pm0.07$ & \nodata & & $26.12\pm0.16$ & \nodata    \\
$ 9$ & $25.91\pm0.17$ & \nodata & & $25.08\pm0.08$ & \nodata & & $25.53\pm0.10$ & \nodata & & $25.01\pm0.09$ & \nodata    \\
$10$ & $24.81\pm0.06$ & $24.13\pm0.08$ & & $25.54\pm0.09$ & $24.57\pm0.26$ & & $25.44\pm0.08$ & $24.68\pm0.11$ & & $25.24\pm0.10$ & $24.39\pm0.09$    \\
$11$ & $25.13\pm0.08$ & \nodata & & $26.09\pm0.15$ & \nodata & & $25.20\pm0.07$ & \nodata & & $25.94\pm0.19$ & \nodata    \\
$12$ & $25.56\pm0.11$ & $24.49\pm0.12$ & & $24.86\pm0.06$ & $24.18\pm0.18$ & & $25.62\pm0.22$ & $24.93\pm0.15$ & & $25.88\pm0.14$ & $24.75\pm0.11$    \\
\hline\\
\colhead{} &
\multicolumn{2}{c}{14256}  &
\colhead{} &
\multicolumn{2}{c}{14575}  &
\colhead{} &
\multicolumn{2}{c}{34707}  &
\colhead{} &
\multicolumn{2}{c}{34852} \\
\cline{2-3} \cline{5-6} \cline{8-9} \cline{11-12}
$ 1$ & $25.58\pm0.08$ & $24.87\pm0.12$ & & $25.40\pm0.08$ & $25.08\pm0.25$ & & $24.35\pm0.04$ & $23.39\pm0.04$ & & $24.30\pm0.04$ & $23.52\pm0.05$    \\
$ 2$ & $25.89\pm0.13$ & $24.75\pm0.09$ & & $26.26\pm0.18$ & $25.22\pm0.27$ & & $24.66\pm0.05$ & $23.54\pm0.05$ & & $24.71\pm0.06$ & $23.49\pm0.05$    \\
$ 3$ & $25.94\pm0.11$ & \nodata & & $25.44\pm0.08$ & \nodata & & $24.97\pm0.06$ & \nodata & & $25.13\pm0.08$ & \nodata    \\
$ 4$ & \nodata & \nodata & & $25.66\pm0.11$ & \nodata & & $24.91\pm0.06$ & \nodata & & $25.16\pm0.07$ & \nodata    \\
$ 5$ & $26.01\pm0.18$ & $25.12\pm0.15$ & & $25.80\pm0.11$ & $24.67\pm0.15$ & & $24.56\pm0.05$ & $23.62\pm0.05$ & & $25.16\pm0.07$ & $24.07\pm0.08$    \\
$ 6$ & $25.13\pm0.06$ & $24.75\pm0.11$ & & $26.17\pm0.19$ & \nodata & & $24.15\pm0.04$ & $23.22\pm0.04$ & & $25.09\pm0.08$ & $24.15\pm0.11$    \\
$ 7$ & $25.32\pm0.06$ & \nodata & & $26.31\pm0.25$ & \nodata & & $23.78\pm0.06$ & \nodata & & $24.56\pm0.05$ & \nodata    \\
$ 8$ & $26.01\pm0.14$ & \nodata & & $25.80\pm0.27$ & \nodata & & $24.06\pm0.03$ & \nodata & & $24.12\pm0.03$ & \nodata    \\
$ 9$ & $25.79\pm0.10$ & \nodata & & $25.67\pm0.11$ & \nodata & & $24.18\pm0.04$ & \nodata & & $24.24\pm0.04$ & \nodata    \\
$10$ & $25.33\pm0.07$ & $24.71\pm0.11$ & & $26.32\pm0.17$ & $24.84\pm0.17$ & & $24.31\pm0.04$ & $23.29\pm0.04$ & & $24.51\pm0.06$ & $23.48\pm0.04$    \\
$11$ & $25.97\pm0.11$ & \nodata & & $26.27\pm0.16$ & \nodata & & $24.40\pm0.04$ & \nodata & & $24.62\pm0.07$ & \nodata    \\
$12$ & $25.16\pm0.06$ & $24.52\pm0.08$ & & $25.47\pm0.08$ & $24.59\pm0.14$ & & $24.76\pm0.05$ & $23.54\pm0.05$ & & $25.05\pm0.06$ & $23.85\pm0.08$    \\
\hline\\
\\
\\
\\
\\
\colhead{} &
\multicolumn{2}{c}{35053}  &
\colhead{} &
\multicolumn{2}{c}{35297}  &
\colhead{} &
\multicolumn{2}{c}{35576}  &
\colhead{} &
\multicolumn{2}{c}{35613} \\
\cline{2-3} \cline{5-6} \cline{8-9} \cline{11-12}
$ 1$ & $25.55\pm0.12$ & $24.06\pm0.08$ & & $25.35\pm0.20$ & $24.13\pm0.08$ & & $25.56\pm0.32$ & $24.51\pm0.14$ & & $25.17\pm0.11$ & $24.52\pm0.14$    \\
$ 2$ & $25.27\pm0.09$ & $24.01\pm0.07$ & & $25.37\pm0.15$ & $24.31\pm0.10$ & & $25.52\pm0.15$ & $24.42\pm0.13$ & & $25.62\pm0.13$ & $24.65\pm0.19$    \\
$ 3$ & $24.72\pm0.06$ & \nodata & & $24.94\pm0.06$ & \nodata & & $25.26\pm0.09$ & \nodata & & $25.34\pm0.09$ & \nodata    \\
$ 4$ & $24.78\pm0.07$ & \nodata & & $25.09\pm0.07$ & \nodata & & $25.36\pm0.09$ & \nodata & & $25.20\pm0.08$ & \nodata    \\
$ 5$ & $25.02\pm0.10$ & $23.78\pm0.07$ & & $25.27\pm0.09$ & $24.29\pm0.08$ & & $25.89\pm0.14$ & $24.07\pm0.14$ & & $25.36\pm0.09$ & $24.39\pm0.10$    \\
$ 6$ & $25.11\pm0.08$ & $23.81\pm0.06$ & & $25.36\pm0.09$ & $24.36\pm0.09$ & & $25.78\pm0.13$ & $24.64\pm0.15$ & & $25.53\pm0.14$ & $24.77\pm0.14$    \\
$ 7$ & $25.19\pm0.09$ & \nodata & & $25.86\pm0.12$ & \nodata & & $25.50\pm0.12$ & \nodata & & $25.40\pm0.10$ & \nodata    \\
$ 8$ & $25.31\pm0.10$ & \nodata & & $24.72\pm0.05$ & \nodata & & $24.63\pm0.05$ & \nodata & & $24.70\pm0.06$ & \nodata    \\
$ 9$ & $25.14\pm0.09$ & \nodata & & $24.78\pm0.05$ & \nodata & & $25.07\pm0.08$ & \nodata & & $25.25\pm0.09$ & \nodata    \\
$10$ & $24.43\pm0.05$ & $23.45\pm0.05$ & & $25.45\pm0.12$ & $24.24\pm0.08$ & & $25.70\pm0.14$ & $24.48\pm0.12$ & & $25.70\pm0.13$ & $24.85\pm0.24$    \\
$11$ & $24.82\pm0.07$ & \nodata & & $25.75\pm0.12$ & \nodata & & $25.59\pm0.10$ & \nodata & & $25.70\pm0.12$ & \nodata    \\
$12$ & $25.24\pm0.09$ & $23.99\pm0.08$ & & $24.52\pm0.04$ & $23.72\pm0.12$ & & $24.51\pm0.13$ & $23.93\pm0.06$ & & $24.59\pm0.27$ & $24.17\pm0.08$    \\
\hline\\
\colhead{} &
\multicolumn{2}{c}{35713}  &
\colhead{} &
\multicolumn{2}{c}{35797}  &
\colhead{} &
\multicolumn{2}{c}{35834}  &
\colhead{} &
\multicolumn{2}{c}{35875} \\
\cline{2-3} \cline{5-6} \cline{8-9} \cline{11-12}
$ 1$ & $26.27\pm0.26$ & $24.74\pm0.33$ & & $25.67\pm0.14$ & $25.46\pm0.34$ & & $25.24\pm0.14$ & $24.46\pm0.13$ & & $26.15\pm0.19$ & $25.14\pm0.20$    \\
$ 2$ & $25.39\pm0.13$ & $24.59\pm0.35$ & & $25.01\pm0.08$ & $23.88\pm0.11$ & & $25.80\pm0.15$ & $24.61\pm0.31$ & & $25.21\pm0.12$ & $24.52\pm0.08$    \\
$ 3$ & $25.69\pm0.17$ & \nodata & & $25.64\pm0.13$ & \nodata & & $25.09\pm0.08$ & \nodata & & $25.94\pm0.12$ & \nodata    \\
$ 4$ & $25.49\pm0.11$ & \nodata & & $25.94\pm0.16$ & \nodata & & $25.33\pm0.13$ & \nodata & & $25.05\pm0.10$ & \nodata    \\
$ 5$ & $24.74\pm0.10$ & $24.31\pm0.10$ & & $25.91\pm0.16$ & $24.97\pm0.21$ & & $25.88\pm0.14$ & $24.49\pm0.15$ & & $25.23\pm0.08$ & $24.44\pm0.08$    \\
$ 6$ & $25.41\pm0.14$ & $24.42\pm0.11$ & & $25.05\pm0.08$ & $24.40\pm0.10$ & & $25.84\pm0.15$ & $24.73\pm0.16$ & & $25.50\pm0.11$ & $24.70\pm0.10$    \\
$ 7$ & $25.73\pm0.13$ & \nodata & & $24.87\pm0.07$ & \nodata & & $25.79\pm0.19$ & \nodata & & $25.83\pm0.25$ & \nodata    \\
$ 8$ & $25.65\pm0.13$ & \nodata & & $25.34\pm0.10$ & \nodata & & $25.02\pm0.08$ & \nodata & & $25.10\pm0.10$ & \nodata    \\
$ 9$ & $24.83\pm0.07$ & \nodata & & $25.74\pm0.17$ & \nodata & & $25.28\pm0.08$ & \nodata & & $25.31\pm0.08$ & \nodata    \\
$10$ & $25.61\pm0.13$ & $24.44\pm0.11$ & & $25.70\pm0.16$ & $25.46\pm0.37$ & & $25.99\pm0.19$ & $24.72\pm0.16$ & & $25.88\pm0.13$ & $24.70\pm0.17$    \\
$11$ & $25.71\pm0.14$ & \nodata & & $25.01\pm0.07$ & \nodata & & $25.75\pm0.25$ & \nodata & & $24.90\pm0.06$ & \nodata    \\
$12$ & $25.06\pm0.08$ & $24.27\pm0.10$ & & $25.58\pm0.12$ & $25.05\pm0.18$ & & $25.38\pm0.09$ & $24.39\pm0.09$ & & $25.64\pm0.16$ & $24.93\pm0.16$    \\
\hline\\
\colhead{} &
\multicolumn{2}{c}{36040}  &
\colhead{} &
\multicolumn{2}{c}{54700}  &
\colhead{} &
\multicolumn{2}{c}{55146}  &
\colhead{} &
\multicolumn{2}{c}{55599} \\
\cline{2-3} \cline{5-6} \cline{8-9} \cline{11-12}
$ 1$ & $25.88\pm0.16$ & $24.71\pm0.18$ & & $24.49\pm0.05$ & $23.43\pm0.05$ & & $24.74\pm0.11$ & $23.85\pm0.06$ & & $24.95\pm0.06$ & $24.00\pm0.07$    \\
$ 2$ & $26.06\pm0.20$ & $25.08\pm0.22$ & & $24.36\pm0.11$ & $23.53\pm0.05$ & & $25.12\pm0.11$ & $24.28\pm0.10$ & & $25.28\pm0.08$ & $24.34\pm0.09$    \\
$ 3$ & $25.30\pm0.09$ & \nodata & & $23.89\pm0.03$ & \nodata & & $25.34\pm0.09$ & \nodata & & $25.50\pm0.12$ & \nodata    \\
$ 4$ & $25.36\pm0.09$ & \nodata & & $23.83\pm0.07$ & \nodata & & $25.20\pm0.08$ & \nodata & & $25.73\pm0.12$ & \nodata    \\
$ 5$ & $25.62\pm0.11$ & $24.57\pm0.12$ & & $23.88\pm0.04$ & $23.09\pm0.03$ & & $24.40\pm0.04$ & $23.61\pm0.05$ & & $25.76\pm0.16$ & $24.50\pm0.10$    \\
$ 6$ & $25.50\pm0.24$ & $24.51\pm0.14$ & & $23.99\pm0.03$ & $23.08\pm0.03$ & & $24.52\pm0.05$ & $23.67\pm0.05$ & & $24.71\pm0.05$ & $23.86\pm0.07$    \\
$ 7$ & $26.00\pm0.17$ & \nodata & & $24.05\pm0.04$ & \nodata & & $24.55\pm0.05$ & \nodata & & $24.67\pm0.05$ & \nodata    \\
$ 8$ & $26.02\pm0.16$ & \nodata & & $24.07\pm0.05$ & \nodata & & $24.92\pm0.06$ & \nodata & & $24.93\pm0.06$ & \nodata    \\
$ 9$ & $26.11\pm0.25$ & \nodata & & $24.23\pm0.04$ & \nodata & & $25.13\pm0.07$ & \nodata & & $25.08\pm0.06$ & \nodata    \\
$10$ & $25.09\pm0.07$ & $24.09\pm0.10$ & & $24.42\pm0.05$ & $23.32\pm0.04$ & & $25.47\pm0.12$ & $23.75\pm0.16$ & & $25.49\pm0.09$ & $24.41\pm0.09$    \\
$11$ & $25.49\pm0.10$ & \nodata & & $24.38\pm0.05$ & \nodata & & $25.28\pm0.15$ & \nodata & & $25.67\pm0.13$ & \nodata    \\
$12$ & $25.73\pm0.12$ & $24.62\pm0.14$ & & $24.54\pm0.10$ & $23.49\pm0.05$ & & $24.46\pm0.04$ & $23.62\pm0.05$ & & $25.24\pm0.07$ & $24.04\pm0.06$    \\
\hline\\
\colhead{} &
\multicolumn{2}{c}{55605}  &
\colhead{} &
\multicolumn{2}{c}{55614}  &
\colhead{} &
\multicolumn{2}{c}{55961}  &
\colhead{} &
\multicolumn{2}{c}{56190} \\
\cline{2-3} \cline{5-6} \cline{8-9} \cline{11-12}
$ 1$ & $24.82\pm0.07$ & $24.06\pm0.07$ & & $25.66\pm0.14$ & $24.60\pm0.13$ & & $25.58\pm0.11$ & $24.45\pm0.11$ & & $25.06\pm0.07$ & $24.42\pm0.13$    \\
$ 2$ & $25.28\pm0.12$ & $24.19\pm0.09$ & & $25.36\pm0.08$ & $24.31\pm0.08$ & & $26.43\pm0.26$ & $24.46\pm0.12$ & & $26.04\pm0.17$ & $24.56\pm0.18$    \\
$ 3$ & $25.60\pm0.11$ & \nodata & & $25.88\pm0.13$ & \nodata & & $25.03\pm0.27$ & \nodata & & $25.27\pm0.08$ & \nodata    \\
$ 4$ & $25.56\pm0.13$ & \nodata & & $24.63\pm0.05$ & \nodata & & $24.96\pm0.07$ & \nodata & & $25.76\pm0.12$ & \nodata    \\
$ 5$ & $25.64\pm0.11$ & $24.20\pm0.14$ & & $24.75\pm0.05$ & $23.90\pm0.13$ & & $25.09\pm0.08$ & $23.91\pm0.06$ & & $25.74\pm0.11$ & $24.63\pm0.20$    \\
$ 6$ & $24.74\pm0.05$ & $23.97\pm0.06$ & & $25.22\pm0.09$ & $24.20\pm0.09$ & & $25.06\pm0.07$ & $23.94\pm0.06$ & & $25.72\pm0.11$ & $24.51\pm0.15$    \\
$ 7$ & $24.75\pm0.06$ & \nodata & & $25.66\pm0.11$ & \nodata & & $25.47\pm0.12$ & \nodata & & $25.06\pm0.06$ & \nodata    \\
$ 8$ & $25.06\pm0.07$ & \nodata & & $25.72\pm0.11$ & \nodata & & $25.40\pm0.11$ & \nodata & & $25.52\pm0.10$ & \nodata    \\
$ 9$ & $25.34\pm0.11$ & \nodata & & $24.69\pm0.05$ & \nodata & & $25.45\pm0.09$ & \nodata & & $25.79\pm0.15$ & \nodata    \\
$10$ & $26.28\pm0.32$ & $24.59\pm0.36$ & & $25.34\pm0.08$ & $24.09\pm0.09$ & & $25.82\pm0.14$ & $24.25\pm0.09$ & & $25.17\pm0.09$ & $24.18\pm0.12$    \\
$11$ & $25.79\pm0.14$ & \nodata & & $25.56\pm0.10$ & \nodata & & $25.69\pm0.13$ & \nodata & & $25.42\pm0.09$ & \nodata    \\
$12$ & $24.92\pm0.07$ & $24.08\pm0.08$ & & $24.72\pm0.05$ & $23.94\pm0.06$ & & $25.93\pm0.14$ & $24.67\pm0.14$ & & $25.63\pm0.10$ & $24.57\pm0.16$    \\
\hline\\
\colhead{} &
\multicolumn{2}{c}{56430}  &
\colhead{} &
\multicolumn{2}{c}{56762}  &
\colhead{} &
\multicolumn{2}{c}{56966}  &
\colhead{} &
\multicolumn{2}{c}{57127} \\
\cline{2-3} \cline{5-6} \cline{8-9} \cline{11-12}
$ 1$ & $25.37\pm0.11$ & $24.55\pm0.13$ & & $25.80\pm0.12$ & $24.79\pm0.12$ & & $26.20\pm0.25$ & $25.03\pm0.16$ & & $25.11\pm0.10$ & $24.55\pm0.10$    \\
$ 2$ & $25.97\pm0.18$ & $24.62\pm0.22$ & & $25.15\pm0.07$ & $24.36\pm0.16$ & & $25.12\pm0.10$ & $24.48\pm0.12$ & & $26.38\pm0.21$ & $25.08\pm0.16$    \\
$ 3$ & $25.48\pm0.09$ & \nodata & & $26.02\pm0.14$ & \nodata & & $26.24\pm0.25$ & \nodata & & $25.97\pm0.15$ & \nodata    \\
$ 4$ & $25.62\pm0.10$ & \nodata & & $26.02\pm0.14$ & \nodata & & $26.06\pm0.16$ & \nodata & & $26.01\pm0.19$ & \nodata    \\
$ 5$ & $25.77\pm0.12$ & $24.71\pm0.16$ & & $26.36\pm0.18$ & $24.86\pm0.12$ & & $26.34\pm0.19$ & $25.20\pm0.19$ & & $26.63\pm0.24$ & $25.03\pm0.15$    \\
$ 6$ & $26.17\pm0.45$ & $24.70\pm0.15$ & & $25.07\pm0.07$ & $24.30\pm0.07$ & & $25.19\pm0.09$ & \nodata & & $26.09\pm0.32$ & $24.84\pm0.12$    \\
$ 7$ & $25.81\pm0.14$ & \nodata & & $25.38\pm0.12$ & \nodata & & $24.92\pm0.21$ & \nodata & & $25.31\pm0.30$ & \nodata    \\
$ 8$ & $24.84\pm0.08$ & \nodata & & $25.57\pm0.10$ & \nodata & & $25.53\pm0.12$ & \nodata & & $26.07\pm0.19$ & \nodata    \\
$ 9$ & $25.51\pm0.11$ & \nodata & & $25.99\pm0.13$ & \nodata & & $25.73\pm0.19$ & \nodata & & $26.33\pm0.18$ & \nodata    \\
$10$ & $26.09\pm0.17$ & $24.81\pm0.15$ & & $25.97\pm0.16$ & $24.95\pm0.16$ & & $26.06\pm0.18$ & $24.59\pm0.12$ & & $25.52\pm0.12$ & $24.35\pm0.08$    \\
$11$ & $25.27\pm0.08$ & \nodata & & $25.25\pm0.07$ & \nodata & & $25.14\pm0.10$ & \nodata & & $25.82\pm0.14$ & \nodata    \\
$12$ & $25.38\pm0.10$ & $24.31\pm0.13$ & & $25.86\pm0.14$ & $24.44\pm0.12$ & & $25.61\pm0.12$ & $24.59\pm0.16$ & & $25.20\pm0.07$ & $24.37\pm0.11$    \\
\hline\\
\\
\\
\\
\\
\\
\\
\colhead{} &
\multicolumn{2}{c}{58373} \\
\cline{2-3} 
$ 1$ & $25.32\pm0.11$ & $24.52\pm0.25$    \\
$ 2$ & $26.96\pm0.37$ & $25.15\pm0.17$    \\
$ 3$ & $25.43\pm0.17$ & \nodata    \\
$ 4$ & $25.79\pm0.11$ & \nodata    \\
$ 5$ & $26.41\pm0.19$ & $25.23\pm0.24$    \\
$ 6$ & $26.39\pm0.26$ & $25.19\pm0.17$    \\
$ 7$ & $26.51\pm0.26$ & \nodata    \\
$ 8$ & $25.50\pm0.10$ & \nodata    \\
$ 9$ & $26.27\pm0.17$ & \nodata    \\
$10$ & $26.50\pm0.25$ & $25.46\pm0.23$    \\
$11$ & $25.58\pm0.09$ & \nodata    \\
$12$ & $26.58\pm0.23$ & $25.09\pm0.38$    \\

\enddata
\label{tab:tab7}

\end{deluxetable}

\clearpage


%% file: tab8.tex

\clearpage
\begin{deluxetable}{lllccrrrr}
\tabletypesize{\scriptsize}
\tablenum{8}
\tablewidth{0pt}
\tablecaption{Properties of Cepheid Variables in NGC 1637}
\tablehead{\colhead{} &
\colhead{$P$} &
\colhead{} &
\colhead{$<V>$} &
\colhead{$<I>$} &
\colhead{$A_V$} &
\colhead{$\mu_V$} &
\colhead{$\mu_I$} &
\colhead{$\mu_W$} \\
\colhead{ID} &
\colhead{(days)} &
\colhead{$\log P$} &
\colhead{(mag)} &
\colhead{(mag)} &
\colhead{(mag)} &
\colhead{(mag)} &
\colhead{(mag)} &
\colhead{(mag)}}

\startdata
$  163\tablenotemark{a}$ & $35.07$ & $1.54$ & $24.60$ & $23.63$ & $ 0.42$ & $30.32$ & $30.15$ & $29.896$ \\
$  282\tablenotemark{a}$ & $35.59$ & $1.55$ & $25.00$ & $23.96$ & $ 0.59$ & $30.74$ & $30.50$ & $30.148$ \\
$  338\tablenotemark{a}$ & $33.49$ & $1.52$ & $25.14$ & $24.15$ & $ 0.50$ & $30.81$ & $30.60$ & $30.310$ \\
$  388\tablenotemark{a}$ & $36.57$ & $1.56$ & $25.19$ & $24.18$ & $ 0.52$ & $30.96$ & $30.75$ & $30.439$ \\
$  540\tablenotemark{a}$ & $24.13$ & $1.38$ & $25.49$ & $24.66$ & $ 0.16$ & $30.76$ & $30.69$ & $30.596$ \\
$  614                 $ & $13.87$ & $1.14$ & $25.57$ & $24.81$ & $ 0.12$ & $30.18$ & $30.13$ & $30.057$ \\
$  706                 $ & $19.54$ & $1.29$ & $25.79$ & $24.81$ & $ 0.57$ & $30.81$ & $30.58$ & $30.241$ \\
$  721                 $ & $21.28$ & $1.33$ & $25.70$ & $24.76$ & $ 0.45$ & $30.82$ & $30.64$ & $30.372$ \\
$  834                 $ & $13.57$ & $1.13$ & $25.91$ & $24.88$ & $ 0.78$ & $30.50$ & $30.18$ & $29.721$ \\
$  852                 $ & $12.31$ & $1.09$ & $25.93$ & $24.98$ & $ 0.61$ & $30.40$ & $30.15$ & $29.794$ \\
$13638                 $ & $19.38$ & $1.29$ & $25.28$ & $24.42$ & $ 0.29$ & $30.29$ & $30.17$ & $30.003$ \\
$13651\tablenotemark{a}$ & $28.80$ & $1.46$ & $25.29$ & $24.18$ & $ 0.83$ & $30.78$ & $30.44$ & $29.949$ \\
$13923\tablenotemark{a}$ & $32.84$ & $1.52$ & $25.37$ & $24.40$ & $ 0.45$ & $31.01$ & $30.83$ & $30.565$ \\
$13938                 $ & $21.83$ & $1.34$ & $25.41$ & $24.54$ & $ 0.28$ & $30.56$ & $30.45$ & $30.278$ \\
$14003                 $ & $15.26$ & $1.18$ & $25.34$ & $24.43$ & $ 0.47$ & $30.06$ & $29.87$ & $29.597$ \\
$14171\tablenotemark{a}$ & $26.91$ & $1.43$ & $25.50$ & $24.50$ & $ 0.55$ & $30.90$ & $30.68$ & $30.355$ \\
$14256                 $ & $15.80$ & $1.20$ & $25.62$ & $24.87$ & $ 0.06$ & $30.39$ & $30.36$ & $30.325$ \\
$14575                 $ & $19.47$ & $1.29$ & $25.84$ & $24.85$ & $ 0.60$ & $30.86$ & $30.61$ & $30.260$ \\
$34707\tablenotemark{a}$ & $50.29$ & $1.70$ & $24.38$ & $23.41$ & $ 0.36$ & $30.53$ & $30.39$ & $30.178$ \\
$34852\tablenotemark{a}$ & $42.42$ & $1.63$ & $24.63$ & $23.63$ & $ 0.44$ & $30.58$ & $30.40$ & $30.133$ \\
$35053\tablenotemark{a}$ & $31.46$ & $1.50$ & $24.92$ & $23.76$ & $ 0.91$ & $30.51$ & $30.14$ & $29.602$ \\
$35297                 $ & $22.88$ & $1.36$ & $25.09$ & $24.11$ & $ 0.54$ & $30.30$ & $30.08$ & $29.754$ \\
$35576\tablenotemark{a}$ & $23.15$ & $1.36$ & $25.22$ & $24.20$ & $ 0.63$ & $30.44$ & $30.18$ & $29.809$ \\
$35613                 $ & $21.13$ & $1.32$ & $25.24$ & $24.42$ & $ 0.16$ & $30.35$ & $30.28$ & $30.186$ \\
$35713                 $ & $17.05$ & $1.23$ & $25.39$ & $24.53$ & $ 0.30$ & $30.24$ & $30.12$ & $29.943$ \\
$35797                 $ & $20.66$ & $1.32$ & $25.37$ & $24.58$ & $ 0.11$ & $30.46$ & $30.41$ & $30.347$ \\
$35834                 $ & $19.35$ & $1.29$ & $25.47$ & $24.55$ & $ 0.45$ & $30.48$ & $30.30$ & $30.033$ \\
$35875                 $ & $15.35$ & $1.19$ & $25.41$ & $24.66$ & $ 0.06$ & $30.14$ & $30.12$ & $30.081$ \\
$36040\tablenotemark{a}$ & $31.71$ & $1.50$ & $25.59$ & $24.52$ & $ 0.68$ & $31.19$ & $30.91$ & $30.506$ \\
$54700\tablenotemark{a}$ & $54.42$ & $1.74$ & $24.19$ & $23.25$ & $ 0.27$ & $30.44$ & $30.33$ & $30.171$ \\
$55146\tablenotemark{a}$ & $32.66$ & $1.51$ & $24.91$ & $23.97$ & $ 0.38$ & $30.55$ & $30.39$ & $30.173$ \\
$55599\tablenotemark{a}$ & $33.58$ & $1.53$ & $25.17$ & $24.22$ & $ 0.40$ & $30.84$ & $30.68$ & $30.448$ \\
$55605\tablenotemark{a}$ & $32.57$ & $1.51$ & $25.22$ & $24.23$ & $ 0.49$ & $30.85$ & $30.65$ & $30.364$ \\
$55614                 $ & $19.22$ & $1.28$ & $25.26$ & $24.31$ & $ 0.50$ & $30.26$ & $30.05$ & $29.755$ \\
$55961\tablenotemark{a}$ & $48.85$ & $1.69$ & $25.43$ & $24.15$ & $ 1.12$ & $31.55$ & $31.09$ & $30.433$ \\
$56190                 $ & $15.47$ & $1.19$ & $25.48$ & $24.43$ & $ 0.81$ & $30.22$ & $29.89$ & $29.412$ \\
$56430                 $ & $20.00$ & $1.30$ & $25.48$ & $24.48$ & $ 0.60$ & $30.53$ & $30.28$ & $29.923$ \\
$56762                 $ & $21.10$ & $1.32$ & $25.58$ & $24.53$ & $ 0.74$ & $30.69$ & $30.39$ & $29.952$ \\
$56966                 $ & $22.41$ & $1.35$ & $25.58$ & $24.62$ & $ 0.50$ & $30.76$ & $30.56$ & $30.267$ \\
$57127                 $ & $15.14$ & $1.18$ & $25.76$ & $24.70$ & $ 0.82$ & $30.48$ & $30.14$ & $29.657$ \\
$58373                 $ & $16.90$ & $1.23$ & $25.99$ & $24.96$ & $ 0.73$ & $30.83$ & $30.54$ & $30.104$ \\

\enddata
\tablenotetext{a}{Used in final distance determination.}

\label{tab:tab8}
\end{deluxetable}

\clearpage


%% file: tab9.tex

\clearpage
\begin{deluxetable}{lll}
\tabletypesize{\footnotesize}
\tablenum{9}
\tablewidth{0pt}
\tablecaption{NGC 1637 Distance Estimates}
\tablehead{\colhead{} &
\colhead{Distance} &
\colhead{} \\
\colhead{Technique} &
\colhead{(Mpc)} &
\colhead{References}}

\startdata
EPM       & $7.86 \pm 0.50\ {\rm (statistical)}$ & 1, 2, 3 \\
Cepheids  & $11.71 \pm 0.99$                & 4       \\
$B$-band TF & $13.8 \pm 2.4$                & 5       \\
Inner ring  & $11.12 \pm 3.53$              & 6       \\
Kinematic   & $10.7 \pm 2.4\ {\rm (statistical)}$ & 7      \\
SEAM (preliminary) & $15.8 \pm 5.7$          & 8      \\
SCM                & $13 \pm 4$              & 9      \\
Plateau-tail      & $11.1 \pm 2.2\ {\rm (statistical)}$ & 10 \\
BRSG               & $7.8 \pm 2.1$           & 11  \\

\enddata 

\tablerefs{ (1) \citealt{Hamuy01b}; (2) \citealt{Leonard5}; (3)
\citealt{Elmhamdi03}; (4) this work; (5) \citealt{Bottinelli85}; (6)
\citealt{Buta83}; (7) \citealt{Hamuy03}; (8) E. Baron et al., in preparation; (9) 
\citealt{Hamuy02}; (10) \citealt{Nadyozhin03}; (11) \citealt{Sohn98}.}

\label{tab:tab9}
\end{deluxetable}

\clearpage
